\documentclass[11pt,a4paper]{article}
\usepackage[top=2cm,bottom=3cm,left=2cm,right=2cm]{geometry}
\usepackage{amsmath}
\usepackage{amssymb}
\usepackage{xcolor}
\usepackage{graphicx}
\usepackage{url}
\usepackage[normalem]{ulem}
\usepackage[%
    colorlinks=true,linkcolor=cyan
]{hyperref}
\usepackage[%
    backend=biber,
    sorting=none,
    style=numeric,
    giveninits=true,
    maxbibnames=99
]{biblatex}
\newcommand{\dd}[1]{\mathrm{d}#1\,}

\def\JAN#1{{\textcolor{magenta}{#1}}}

\title{Stochastic interpolation of sparsely sampled time series by a superstatistical random process and its synthesis in Fourier and wavelet space}
\author{Jeremiah Lübke$^{1,}$\thanks{E-mail: \href{mailto:jeremiah.luebke@ruhr-uni-bochum.de}{jeremiah.luebke@ruhr-uni-bochum.de}} \and Jan Friedrich$^2$ \and Rainer Grauer$^1$}
\date{%
\begin{minipage}[t]{.4\textwidth}
    \centering
  $^1$Institute for Theoretical Physics \\ Ruhr-University Bochum \\ Universitätsstr.~150, 44892 Bochum
\end{minipage} \quad
\begin{minipage}[t]{.4\textwidth}
    \centering
  $^2$ForWind, Institute of Physics \\ University of Oldenburg \\ Küpkersweg 70, 26129 Oldenburg
\end{minipage} \\[1em]
    \emph{Date: \today}
}

\begin{document}

\maketitle

\begin{abstract}
    We present a novel method for stochastic interpolation of sparsely sampled time signal\JAN{s} based on a superstatistical random process generated from a multivariate Gaussian scale mixture.
    In comparison to other stochastic interpolation methods such as Gaussian process regression, our method possesses strong multifractal properties and is thus applicable to a broad range of real-world time series, e.g.~from solar wind or atmospheric turbulence.
    Furthermore, we provide a sampling algorithm in terms of a mixing procedure that consists of generating a~$1+1$-dimensional field~$u(t,\xi)$, where each Gaussian component~$u_\xi(t)$ is synthesized with identical underlying noise but different covariance function~$C_\xi(t,s)$ parameterized by a log-normally distributed parameter~$\xi$.
    Due to the Gaussianity of each component~$u_\xi(t)$, we can exploit standard sampling alogrithms such as Fourier or wavelet methods and, most importantly, methods to constrain the process on the sparse measurement points.
    The scale mixture~$u(t)$ is then initialized by assigning each point in time~$t$ a~$\xi(t)$ and therefore a specific value from~$u(t,\xi)$, where the time-dependent parameter~$\xi(t)$ follows a log-normal process with a large correlation time scale compared to the correlation time of~$u(t,\xi)$.
    We juxtapose Fourier and wavelet methods and show that a multiwavelet-based hierarchical approximation of the interpolating paths, which produce a sparse covariance structure, provide an adequate method to locally interpolate large and sparse datasets.
    \vskip\baselineskip\noindent
\end{abstract}

\section{Introduction}
The need to construct realistic synthetic turbulent fields has increased
significantly in recent years in various application environments. Examples
include the influence of turbulent fluctuations on wind turbine loads~\cite{muecke-kleinhans-peinke:2011}, the development of active control
mechanisms for wind farms~\cite{spencer-stol-unsworth-etal:2013} and, in
particular, cosmic ray particle transport in space and astrophysical plasmas~\cite{schlickeiser:2015, bustard-zweibel:2021}.

The most natural approach to the study of charged particle transport in turbulent magnetic fields consists in a numerical solution of the magnetohydrodynamic (MHD) equations and
the propagation of particles in the dynamical turbulent MHD fields. This
approach has been applied  by various authors (see e.g.~\cite{cohet-marcowith:2016}) to simulate small regions of interest in order to extract
local transport processes and effective diffusion coefficients as input for coarse-grained
transport descriptions.
Describing significantly larger regions of
interest (e.g., the heliosphere, see below, the interstellar medium, or galactic
outflows) by this method, however, imposes major computational challenges: for example, in order to simulate the entire heliosphere one has to resolve small
scales at the ion-gyro radius ($\sim10^5$\,m), where damping processes operate,
up to about 100\,AU, which is the distance to the solar wind termination shock.
This would result in a mesh of about~$1,500,000^3$~points, a size which is not
possible to handle in the near future. Therefore, one of the main problems is the
design of efficient sampling methods for synthetic turbulent fields in the limit of unprecedentedly high Reynolds numbers.

In addition, the synthetic fields must be embedded in existing large-scale
MHD-like simulations by an appropriate stochastic interpolation.
Furthermore, the construction should be able to synthesise anisotropy (in the
sense of~\cite{goldreich-sridhar:1995,boldyrev:1995}) and small-scale
intermittency of real turbulent field fluctuations.
Especially the incorporation intermittent fluctuations, which manifest itself by heavy-tailed probability distributions at small scale, imposes major challenges for the modeling procedure as one cannot rely on commonly used Gaussian random field algorithms.

Tremendous work on particle transport in synthesised turbulent fields has been done already, starting with the pioneering work by~\cite{giacalone-jokipii:1999}. This and subsequent work (e.g.~\cite{qin-matthaeus-etal:2002, tautz-dosch:2013,
reichherzer-etal:2020} and references therein) were based on a superposition of
Fourier modes aimed at testing for which regimes the transport remains diffusive
(as opposed to, e.g., subdiffusive perpendicular to the magnetic guide  field).
Most of these simulations made use of specific turbulence spectra like
isotropic, slab, or slab/two-dimensional ones and employed a homogeneous guide
field (see e.g.~\cite{dundovic-pezzi-blasi-etal:2020} where a detailed
comparison of these approaches with respect to parallel and perpendicular
diffusion is presented).

Studies of particle transport in anisotropic turbulence~\cite{pommois-zimbardo-veltri:2007} have already shown that, in addition to
diffusion, superdiffusion as well as subdiffusion can occur subject to the
strength of the anisotropy. 
Several papers address the effects of intermittency on particle transport.
In~\cite{alouanibibi-leroux:2014} an
intermittency model was introduced, in which the deviation from a Gaussian PDF is generated by modifying the amplitudes of the plane wave modes with a~$q$-Gaussian statistic.
%
In~\cite{pucci-malara-etal:2016} the so-called~$p$-model~\cite{meneveau-sreenivasan:1987} was applied to generate intermittent turbulent fields. One conclusion from this study was that intermittency mainly increases parallel transport.
In~\cite{shukurov-snodin-etal:2017} an intermittent magnetic field by solving the induction equation with a given velocity field was introduced. Their conclusion was that intermittency has a profound effect on the diffusion tensor, especially for energies~$E < 10^{10}$\,GeV.
In~\cite{durrive-lesaffre-ferriere:2020} a method for generating magnetic turbulence by generalizing an approach from fluid dynamics~\cite{pereira-garban-chevillard:2016} was presented. This method is based on a generalized Biot-Savart kernel that takes into account the stretching of vorticity encoded in the Cauchy-Green tensor. 

In order to add to this body of work and take first steps in the direction of conditioning intermittent synthetic fields on data obtained by simulations or \emph{in-situ} measurements, we present in this work the construction of a one-dimensional stochastic process, which is suitable for interpolation of coarse data and exhibits intermittency as observed in hydrodynamic turbulence.
The extension to three dimensions, including the divergence-free condition and an anisotropic energy spectrum are prospect for future publications.

Intermittency is accounted for by a superposition or scale mixture of multivariate Gaussians statistics, parametrized by the log-normal model of Kolmogorov and Obukhov~\cite{kolmogorov:1962,obukhov:1962}.
More generally, the idea of superposing Gaussian (or equilibrium) statistics in order to model heavy-tail behavior is referred to as superstatistics~\cite{beck:2003} and has been applied in numerous research areas, as presented in the review article~\cite{beck:2009}.
On the subject of superstatistics and non-Gaussian diffusion, see~\cite{metzler:2020}.
%
For the methodology employed in this paper, we refined the idea first proposed in~\cite{friedrich:2021} and recently applied to the reconstruction of wind fields from point-wise atmospheric turbulence measurements~\cite{friedrich2022arxiv,friedrich2022surrogate}.
This approach essentially constitutes a~$n$-point generalization of classical superstatistics.
A sample drawn from such a superstatistical model consists of Gaussian components,
which can be readily conditioned on coarse data points via a stochastic interpolation equation~\cite{friedrich:2020}.
Finally, an efficient application to large coarse datasets is achieved by means of a hierarchic representation based on a discrete wavelet transformation~\cite{bcr:92,phoon:2004}.
Different approaches based on wavelet transforms for synthesizing Gaussian random fields have been presented in the past in, e.g.,~\cite{zeldin:1996,elliott:1994}.
Intermittent random fields can be synthesized by combining the dyadic structure of the wavelet transform of choice with a multiplicative cascade, such as pioneered in~\cite{benzi:1993}, and later developed into so called~$\mathcal{W}$-cascades~\cite{arneodo:1998}.
See~\cite{muzy:2019} for recent work.

The outline of the paper is as follows: In section~\ref{sec:superstat} we describe the superstatistical model for multifractal random fields according to the log-normal model and introduce the sampling algorithm in section~\ref{sec:algo}.
In section~\ref{sec:fourier} we illustrate the algorithm in one dimension by sampling the Gaussian components with the standard circulant embedding procedure based on the Fourier transform of the covariance function.
These results will serve as a benchmark for the wavelet-based algorithm.
The Fourier-based sampling procedure is mainly limited by the memory consumption when applied to large datasets, which is overcome by formulating the sampling and conditioning procedures in terms of multiwavelets with compact support~\cite{alpert-L2:1993}, which facilitate a more localized construction, as shown in sections~\ref{sec:wavelets} and~\ref{sec:interp}.
Then we empirically verify the statistical properties of the algorithm in the unconstrained and constrained cases in section~\ref{sec:results} and conclude the paper in section~\ref{sec:outlook}.

\section{Superstatistical Model}
\label{sec:superstat}
Turbulence is often studied in terms of the velocity increments over a separation~$\tau$
\begin{equation}
    \delta u_\tau(t)=u(t+\tau)-u(t),
\end{equation}
where we consider a one-dimensional velocity signal~$u(t)$ indexed by time~$t$.
The velocity increments are assumed to be statistically stationary and thus we drop the explicit time dependence.
A key observation is the strong non-Gaussian shape of the increment probability distribution~$p(\delta u_\tau)$, which is characterized by heavy tails for decreasing values of~$\tau$, arising from rare extreme events in the increment time series.
Of particular interest is the scaling behaviour of the structure functions, i.e.~the scale dependent moments of~$p(\delta u_\tau)$
\begin{equation}
    S_p(\tau)=\langle\delta u_\tau^p\rangle\propto\tau^{\zeta_p},
\end{equation}
where intermittency is associated with anomalous scaling of~$S_p(\tau)$, i.e.~$\zeta_p$ being a non-linear function of~$p$.
Almost a century of turbulence research has given raise to an abundance of phenomenological models for~$\zeta_p$, one of the first being the log-normal model by Kolmogorov and Obukhov~\cite{kolmogorov:1962,obukhov:1962}.
While this model has some known shortcomings ($\zeta_p$ must be a non-decreasing function of~$p$ for incompressible flow, which is violated by this model~\cite{frisch:95}), it captivates through its conceptual simplicity and ease of use.
Specifically, the model's prime subject is the energy dissipation rate~$\varepsilon_\tau$ locally averaged over an interval of size~$\tau$, which is modeled as log-normally distributed with~$\log\varepsilon_\tau$ having mean~$\frac{\mu}{2}\log\frac{\tau}{\mathcal{T}}$ and variance~$A+\mu\log\frac{\mathcal{T}}{\tau}$,
where~$\mu$ is the intermittency parameter,~$\mathcal{T}$ is the largest external scale of the system and~$A$ is a constant associated with the macrostructure of the flow.
The scaling behaviour~$\langle\varepsilon_\tau^p\rangle\propto\tau^{\varsigma_p}$ can be related to~$\langle\delta u_\tau^p\rangle\propto\tau^{\zeta_p}$ via Kolmogorov's refined similarity hypothesis as~$\zeta_p=p/3+\varsigma_{p/3}$ leading to
\begin{equation}
    \zeta_p=\frac{p}{3}-\frac{\mu}{18}(p^2-3p).
    \label{eq:lognormal-scaling}
\end{equation}

Beck~\cite{beck:2004} modeled increment distribution~$p(\delta u_\tau)$ as a mixture of Gaussian probability distributions in the framework of \emph{superstatistics}
\begin{equation}
    p(\delta u_\tau)=\sqrt{\frac{\beta}{2\pi}}\int_0^\infty\dd{\beta}f(\beta)\exp\left(-\frac{1}{2}\beta\delta u_\tau^2\right),
\end{equation}
where the signal is assumed to be locally Gaussian with the local variance given by the inverse of an intensive fluctuating parameter~$\beta$ and the global statistics are obtained by averaging over all possible values of~$\beta$.
In the case of turbulence~$\beta$ can be interpreted as a function of the locally averaged energy dissipation rate~$\varepsilon_\tau$, which suggests a log-normal distribution for~$\beta$
\begin{equation}
    f(\beta)=\frac{1}{\beta s\sqrt{2\pi}}\exp\left(-\frac{\left[\log\frac{\beta}{m}\right]^2}{2s^2}\right),
\end{equation}
where~$m$ and~$s$ are parameters to be fitted to data.

\begin{figure}[t]
    \centering
    \includegraphics[width=\textwidth]{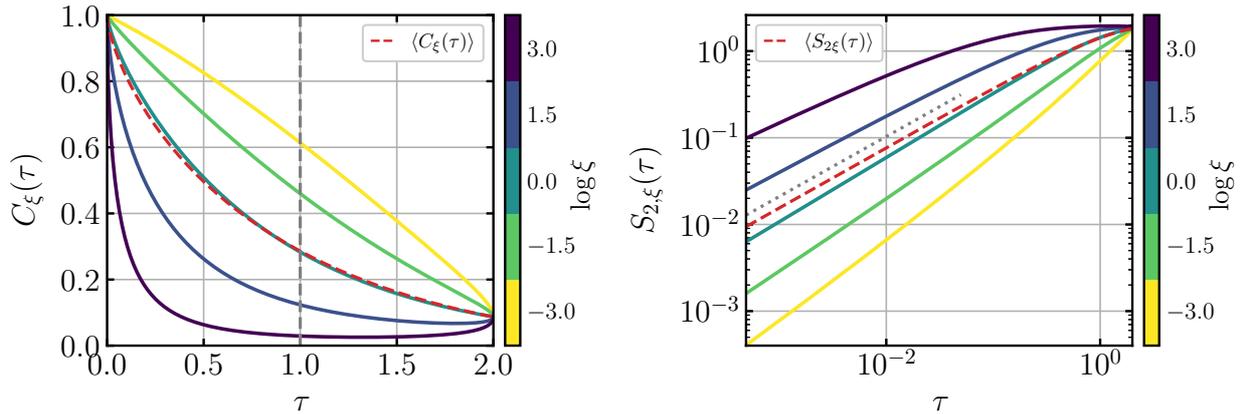}
    \caption{\emph{left:} Illustration of the superstatistical covariance function~$C_\xi(\tau)=C(\varepsilon_{\xi,\tau}\tau)$ for~$\tau\in(0,\mathcal{T})$ and different values of~$\xi$.~$C(\tau)$ is the Matérn covariance function given by equation~(\ref{eq:matern}) with~$H=1/3$ and~$T=1$ (indicated by the vertical dashed line) and the largest length scale of the~$\varepsilon_{\xi,\tau}$-process is~$\mathcal{T}=2$.
    Note that~$C_\xi(\tau)$ is independent of the mixture parameter~$\xi$ at~$\tau=0$ leading to a stationary mixture process, and at~$\tau=\mathcal{T}$ leading to Gaussian statistics on large scales.
    The simulations were performed for smaller scales~$\tau\in(0,T)$.
    Shown is also the covariance function of the mixture process with a red dashed line, obtained by a log-normal weighted average over all~$\xi$, i.e.~$\langle C_\xi(\tau)\rangle=\int_0^\infty\dd{\xi}f(\xi)C_\xi(\tau)$ with~$f(\xi)=(2\pi)^{-1/2}\xi^{-1}\exp\left(-\frac{1}{2}[\log\xi]^2\right)$. \\
    \emph{right:} The second-order structure function for stationary processes is related to the covariance function as~$S_\xi(\tau)=2C_\xi(0)-2C_\xi(\tau)$. A log-log plot of~$S_\xi(\tau)$ emphasizes the behavior on small scales~$\tau\ll T$.
    Shown is again the structure function of the mixture process with a dashed red line, obtained by averaging over all~$\xi$, as well as a dotted grey line with slope~$2/3+\mu/9$, indicating the small-scale scaling of the mixture process.}
    \label{fig:cov}
\end{figure}

Our aim is to  extend this ansatz from the one-point statistics of velocity increments to an~$n$-point description for discretely sampled time series~$\mathbf{u}=(u(t_0),\cdots,u(t_{n-1}))$.
Following Friedrich et al.~\cite{friedrich:2021} we write
\begin{equation}
    p(\mathbf{u})=(2\pi)^{-\frac{n}{2}}\int_{\mathbb{R}^n_+}\dd{\boldsymbol{\xi}}f(\boldsymbol{\xi})|\Sigma_{\boldsymbol\xi}|^{-\frac{1}{2}}\exp\left(-\frac{1}{2}\mathbf{u}^\top\Sigma_{\boldsymbol\xi}^{-1}\mathbf{u}\right),
    \label{eq:multipoint-pdf}
\end{equation}
where the intensive fluctuating parameter~$\boldsymbol\xi$ is now a vector of size~$n$ with real positive entries and probability distribution~$f(\boldsymbol\xi)$ and~$\Sigma_{\boldsymbol\xi}$ is the covariance matrix of the Gaussian components parameterized by~$\boldsymbol\xi$.
The explicit form of~$f(\boldsymbol\xi)$ and the parameterization~$\Sigma_{\boldsymbol\xi}$ will be discussed in the next section.
The Gaussian components of this multipoint mixture model are derived from an underlying process~$\Tilde{u}(t)$ characterized by the unparameterized covariance function~$C(t_j,t_k)=\Sigma_{jk}=\langle\Tilde{u}(t_j)\Tilde{u}(t_k)\rangle$, which defines the one- and two-point behaviour of the mixture process.
Specifically, we want the process to be stationary, i.e.~$C(t_j,t_k)=C(|t_j-t_k|)$, and exhibit fractional scaling with exponent~$H$ in accordance with turbulence scaling laws, i.e.~$S_2(\tau)=2C(0)-2C(\tau)\propto\tau^{2H}$ for small~$\tau$.
A natural way to introduce the superstatistical parameterization is then to stretch the scale~$\tau$ with the locally averaged energy dissipation rate~$\varepsilon_\tau$.
Noting that we can write~$\varepsilon_{\xi,\tau}=\exp\left(\sqrt{A+\mu\log{\mathcal{T}}/{\tau}}\log\xi+\frac{\mu}{2}\log{\tau}/{\mathcal{T}}\right)=\xi^{\sqrt{A+\mu\log{\mathcal{T}}/{\tau}}}\left({\tau}/{\mathcal{T}}\right)^{\mu/2}$ for~$\tau<\mathcal{T}$ and a standard normally distributed~$\log\xi$, we introduce the superstatistical covariance function
\begin{equation}
    C_\xi(\tau)=C(\varepsilon_{\xi,\tau}\, \tau),
    \label{eq:superstat-covariance}
\end{equation}
which is illustrated for different values of~$\xi$ in figure~(\ref{fig:cov}).

We want to emphasize that the intensive fluctuating parameter must be considered as a time-dependent process~$\xi(t)$ which is reflected by the vector-valued parameter~$\boldsymbol\xi$ in equation~(\ref{eq:multipoint-pdf}).
Also,~$\xi(t)$ must vary slowly compared to the underlying process in accordance with the superstatistical assumption of local Gaussian behaviour.
But before we extend equation~(\ref{eq:superstat-covariance}) accordingly, let us verify that our approach indeed recovers the log-normal scaling law equation~(\ref{eq:lognormal-scaling}). 

To this end let~$u(t,\xi(t))$ be a single value from a time series sampled from the mixture model equation~(\ref{eq:multipoint-pdf}), which explicitly depends on the current value of the parameter~$\xi(t)$.
Then the velocity increments over a separation~$\tau$ read~$v=u(t+\tau,\xi(t+\tau))-u(t,\xi(t))$ and we have on average~$\xi(t)\approx\xi(t+\tau)$ for small~$\tau$, since~$\xi$ varies slowly compared to~$u$.
This enables us to write the scale-dependent increment distribution as a Gaussian mixture with a scalar-valued standard log-normally distributed parameter~$\xi$
\begin{equation}
    p(v, \tau)=\int_0^\infty\dd{\xi}f(\xi)\left(2\pi S_{2,\xi}(\tau)\right)^{-1/2}\exp\left(-\frac{v^2}{2S_{2,\xi}(\tau)}\right)
    \label{eq:increment-distribution-ensemble}
\end{equation}
with~$f(\xi)=(2\pi)^{-1/2}\xi^{-1}\exp\left(-\frac{1}{2}[\log\xi]^2\right)$ and variance~$S_{2,\xi}(\tau)=2C_{\xi}(0)-2C_\xi(\tau)\propto(\varepsilon_{\xi,\tau}\tau)^{2H}$.
With this expression, the structure functions can be explicitly computed (see appendix~\ref{app:moments})
\begin{equation}
    S_p(\tau)=\int_{\mathbb{R}}\dd{v}v^p p(v,\tau)=C_p\tau^{pH-\frac{1}{2}\mu H^2(p^2-p/H)},
    \label{eq:structure-functions}
\end{equation}
and for~$H=1/3$ we indeed recover the log-normal scaling law given by equation~(\ref{eq:lognormal-scaling}).


Finally we note, that~$\xi(t)$ (if chosen appropriately) eventually explores the entire interval~$(0,\infty)$, where each value is visited with probability~$f(\xi)$.
This implies that the time-averaged increment distribution of a single time series
\begin{equation}
    p_T(v,\tau)=\frac{1}{T}\int_0^T\dd{t}\delta\big(v-\big[u(t+\tau,\xi(t+\tau))-u(t,\xi(t))\big]\big)
\end{equation}
almost surely converges for large times~$T$ to the ensemble average~$p(v,\tau)$ given by equation~(\ref{eq:increment-distribution-ensemble}), i.e.
\begin{equation}
    \lim_{T\to\infty}p_T(v,\tau)=p(v,\tau).
\end{equation}
Consequently, individual time series described by the mixture model equation~(\ref{eq:multipoint-pdf}) exhibit log-normal intermittency.

Concerning the underlying stationary covariance function~$C(\tau)$, there is some freedom of choice, as long as the small-$\tau$ asymptotics~$C(0)-C(\tau)\propto\tau^{2H}$ are fulfilled.
For example, Friedrich et al.~\cite{friedrich:2021} employed~$C(|t-s|)=\langle w(t)w(s)\rangle$, where~$w(t)$ is an Ornstein-Uhlenbeck process driven by fractional Brownian motion, i.e.~$\dd{w}_t=-\frac{1}{T}w_t\dd{t}+\dd{B^H_t}$ with~$\langle B^H_tB^H_s\rangle=\frac{\sigma^2}{2}\big(|t|^{2H}+|s|^{2H}-|t-s|^{2H}\big)$.
Alternatively one can consider a stochastic process with power spectrum~$S(\omega)\propto{T^{-2H}}{\left(T^{-2}+\omega^2\right)^{-H-1/2}}$ whose Fourier transform gives the \emph{Matérn} covariance function~\cite{lilly:2017}
\begin{equation}
    C(\tau)=\sigma^2\frac{2^{1-H}}{\Gamma(H)}\left(\frac{\tau}{T}\right)^HK_H\left(\frac{\tau}{T}\right),
    \label{eq:matern}
\end{equation}
for~$\tau>0$ where~$\Gamma(z)=\int_0^\infty t^{z-1}e^{-t}\dd{t}$ for~$\mathrm{Re}(z)>0$ denotes the gamma function and~$K_H(z)=\int_0^\infty e^{-z\cosh(t)}\cosh(Ht)\dd{t}$ for~$\mathrm{Re}(z)>0$ denotes the modified Bessel function of the second kind.
For~$H=\frac{1}{2}$ the regular Ornstein-Uhlenbeck process is recovered.
%
The parameter~$T$ in both approaches denotes the characteristic time scale of the process beyond which correlations are exponentially damped.

\section{Sampling Algorithm}
\label{sec:algo}
The standard way to sample from Gaussian mixture models such as equation~(\ref{eq:multipoint-pdf}) reads:
\begin{equation}
    \mathbf{u}=\Sigma_{\boldsymbol\xi}^{1/2}\mathbf{y},
    \label{eq:formal-sampling}
\end{equation}
where~$\boldsymbol\xi=(\xi(t_0),\cdots,\xi(t_{n-1}))\sim f(\boldsymbol\xi)$ denotes the discretely sampled parameter process~$\xi(t)$,~$\mathbf{y}$ a Gaussian white noise vector and~$\Sigma_{\boldsymbol\xi}^{1/2}$ the square root of the parameterized covariance matrix.
We know that the one-point statistics of the logarithm of the parameter process~$\log\xi(t)$ is given by a standard normal distribution, which suggests to model~$\log\boldsymbol\xi$ as a discrete sample from a Gaussian process with zero mean, unit variance and covariance function~$C_{\text{parameter}}(t_j,t_k)$ with a sufficiently large correlation time scale~$T_{\text{parameter}}>T$.
The expression for the probability distribution of~$\boldsymbol\xi$ then reads
\begin{equation}
    f(\boldsymbol\xi)=(2\pi)^{-n/2}|\Theta|^{-1/2}\left(\prod_{j=1}^{n}\xi_j^{-1}\right)\exp\left(-\frac{1}{2}\log\boldsymbol\xi^\top\Theta^{-1}\log\boldsymbol\xi\right),
    \label{eq:parameter-process-distribution}
\end{equation}
with covariance matrix~$\Theta_{jk}=C_{\text{parameter}}(t_j,t_k)$.
In this paper we chose an Ornstein-Uhlenbeck process with~$C_{\text{parameter}}(t_j,t_k)=\exp\left(-|t_j-t_k|/T_{\text{parameter}}\right)$ for the sake of simplicity.

In order to take the explicit dependence of~$u_j=u(t_j,\xi(t_j))$ on~$\xi(t_j)$ into account, we write the~$j$-th row of equation~(\ref{eq:formal-sampling}) as the dot product
\begin{equation}
    u_j=\boldsymbol\sigma_{j,\xi_j}^\top\mathbf{y},
    \label{eq:rowwise-sampling}
\end{equation}
where~$\boldsymbol\sigma_{j,\xi_j}^\top$ is the~$j$-th row of~$\Sigma_{\xi_j}^{1/2}$, i.e.~the square root of the covariance matrix with the scalar-valued parameterization~$\xi_j=\xi(t_j)$.
We can then employ the superstatistical covariance function equation~(\ref{eq:superstat-covariance}) to define the entries of the covariance matrix
\begin{equation}
    \Sigma_{\xi_j,kl}=C_{\xi_j}(|t_k-t_l|).
    \label{eq:superstat-covariance-entries}
\end{equation}
Note that in principle each~$u_j$ is computed with a different~$\Sigma_{\xi_j}$ but with identical noise~$\mathbf{y}$.
To summarize, the vector-valued parameterized covariance matrix can be expressed as
\begin{equation}
    \Sigma_{\boldsymbol\xi}={\Sigma_{\boldsymbol\xi}^{1/2}}{\Sigma_{\boldsymbol\xi}^{1/2}}^\top
    \quad\text{with}\quad
    \Sigma_{\boldsymbol\xi}^{1/2}=\begin{pmatrix}
        \text{---}&\boldsymbol\sigma_{0,\xi_0}&\text{---} \\
        &\vdots& \\
        \text{---}&\boldsymbol\sigma_{n-1,\xi_{n-1}}&\text{---}
    \end{pmatrix}.
    \label{eq:superstat-covariance-matrix}
\end{equation}

We can obtain a numerically efficient approximation of equation~(\ref{eq:rowwise-sampling}) by discretizing~$\xi$-space into~$m$ segments~$\hat{\xi}_0<\cdots<\hat{\xi}_{m-1}$ and considering the~$m\times n$ matrix
\begin{equation}
    U=\begin{pmatrix}
    \text{---}&\mathbf{u}_{\hat{\xi}_0}&\text{---} \\
    &\vdots&\\
    \text{---}&\mathbf{u}_{\hat{\xi}_{m-1}}&\text{---} \\
    \end{pmatrix},
    \label{eq:Umatrix}
\end{equation}
where the rows are Gaussian sample paths ~$\mathbf{u}_{\hat{\xi}_j}=\Sigma_{\hat{\xi}_j}^{1/2}\mathbf{y}$ of length~$n$ with fixed~$\hat{\xi}_j$, again using identical noise~$\mathbf{y}$ for all~$j$.
We then translate the parameter sample path~$\boldsymbol\xi=(\xi_0,\cdots,\xi_{n-1})\sim f(\boldsymbol\xi)$ into an index process
\begin{equation}
    \Xi_j=\mathrm{argmin}_{k=0,\cdots,m-1}|\xi_j-\hat{\xi}_k|
    \label{eq:choice-process}
\end{equation}
and assemble the final mixture sample~$\mathbf{u}$ by element-wise assignment~$u_j\gets U_{\Xi_jj}=\big(\Sigma^{1/2}_{\hat{j}}\mathbf{y}\big)_j$ with
{\vbox to 0pt{\vss\hbox{~$\hat{j}=\hat{\xi}_{\Xi_j}$
}}}, which corresponds to equation~(\ref{eq:rowwise-sampling}).
From a continuous viewpoint the matrix~$U$ is a discrete sample of a~$1+1$-dimensional process~$u(t,\xi)$, where slices~$u(t,\xi')$ with fixed~$\xi'$ are samples of Gaussian processes with covariance functions~$C_{\xi'}(\tau)$ and identical underlying noise.
Given a sample of the parameter process~$\xi(t)$, the assignment for the mixture process reads~$u(t)\gets u(t,\xi(t))$.

\begin{figure}[t]
    \centering
    \includegraphics[width=.6\textwidth]{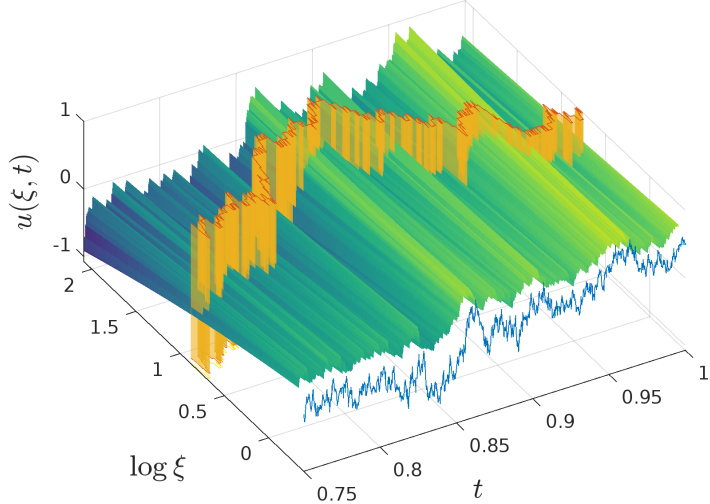}
    \caption{Shown is an excerpt of the mixture construction. The sample matrix~$U$ is plotted as a surface over the discretized~$\log\xi$-$t$ plane, where each row is a Gaussian process realization with covariance~$C_{\xi_i}(\tau)$.
    The choice process~$\Xi(t)$ according to equation~(\ref{eq:choice-process}) is plotted as a vertical surface cutting through the~$U$-plane.
    The point-wise assignment~$u(t)\gets u(t,\xi(t))$ is here represented by projecting the intersection of the~$\Xi(t)$- and~$U$-plane out of~$\log\xi$-space, resulting in the mixture realization~$u(t)$ plotted in front of the surface.}
    \label{fig:mixture}
\end{figure}

Employing equations~(\ref{eq:formal-sampling}) and~(\ref{eq:rowwise-sampling}) directly requires potentially up to~$n$ matrix square root decompositions (assuming all~$\xi_j$'s are different), where only a small number of the resulting rows is actually used.
On the other hand, a discretization of~$\xi$-space in~$m\ll n$ segments makes the procedure much more efficient and enables the full computation of the matrix~$U$, as illustrated in figure~(\ref{fig:mixture}).
This is the preferred approach when the square root of the covariance matrix is not explicitly constructed (as is the case with the Fourier- and Wavelet-based methods discussed below), in which case~$\Sigma_{\boldsymbol\xi}^{1/2}$ according to equation~(\ref{eq:superstat-covariance-matrix}) cannot be directly assembled.
Throughout this work we use~$m=100$ and~$\log\hat{\xi}_j=-3+6j/(m-1)$, such that~$\log\hat{\xi}_0=-3$ and~$\log\hat{\xi}_{m-1}=3$.

\section{Fourier algorithm}
\label{sec:fourier}
Sampling the approximation of the mixture process~$\mathbf{u}$ given by equations~(\ref{eq:Umatrix}) and~(\ref{eq:choice-process}) can be done by sampling~$m$ Gaussian processes~$\mathbf{u}_{\hat{\xi}_k}$ with identical noise and covariance matrices~$\Sigma_{\hat{\xi}_k}$ according to equation~(\ref{eq:superstat-covariance-entries}) for fixed~$\hat{\xi}_0<\cdots<\hat{\xi}_{m-1}$, and sampling the log-normal parameter process~$\boldsymbol\xi\sim f(\boldsymbol\xi)$ according to equation~(\ref{eq:parameter-process-distribution}).
Then each point of the mixture sample gets assigned the corresponding value~$u_j\gets u_{\hat{j},j}$ from the Gaussian process sample labeled by~$\hat{j}=\hat{\xi}_{\Xi_j}$ with~$\Xi_j=\mathrm{argmin}_{k=0,\cdots,m-1}|\xi_j-\hat{\xi}_k|$.

In the case of stationary covariance and uniform grid points, we conveniently perform the sampling of Gaussian processes in Fourier space~\cite{dietrich:93}.
Generally, a Gaussian process sample with covariance matrix~$\Sigma$ and noise vector~$\mathbf{y}$ is given by~$\mathbf{u}=\Sigma^{1/2}\mathbf{y}$, where the matrix square root may be expressed as~$\Sigma^{1/2}=F^\dag\Lambda^{1/2}$ given the eigen-decomposition~$\Sigma=F^\dag\Lambda F$, where the diagonal matrix~$\Lambda_{jj}=\lambda_j$ contains the eigenvalues of~$\Sigma$ and~$F^\dag$ denotes the conjugate transpose of~$F$.
The expression~$\Lambda^{1/2}_{jj}=\sqrt{\lambda_j}$ is well defined because covariance matrices are by definition positive semi-definite, i.e. they only have non-negative eigenvalues.
If~$\Sigma$ is symmetric, periodic and stationary, i.e.~its first row has the form~$\mathbf{c}=(c_0,\cdots,c_n,c_{n-1},\cdots,c_1)\in\mathbb{R}^{2n}$ and all its diagonals are constant, then~$\Sigma$ is diagonalized by the discrete Fourier transform with~$F_{jk}=\frac{1}{\sqrt{2n}}e^{2\pi\mathrm{i}jk/2n}$ for~$j,k=0,\cdots,2n-1$ and its eigenvalues are given by~$\Lambda=\mathrm{diag}(F\mathbf{c})$.
Two independent samples of the Gaussian process are then given by the the real and imaginary parts of~$\mathbf{u}=F^\dag\Lambda^{1/2}\mathbf{y}$ with complex-valued Gaussian white noise~$\mathbf{y}=\frac{1}{\sqrt{2}}(\mathbf{y}_{\mathrm{re}}+\mathrm{i}\mathbf{y}_{\mathrm{im}})$.
Matrix-vector products with the Fourier matrices~$F$ and~$F^\dag$ are efficiently evaluated with the FFT and iFFT algorithms respectively.
To summarize, if the superstatistical covariance functions~$C_\xi(\tau)$ are periodic on~$\tau\in[0,2]$ with the discretization~$\tau_k=2k/(2n-1)$ for~$k=0,\cdots,2n-1$, the mixture process equation~(\ref{eq:formal-sampling}) reads in Fourier space:
\begin{subequations}
\begin{gather}
    u_{\mathrm{re},j}+\mathrm{i}u_{\mathrm{im},j}=\frac{1}{\sqrt{4n}}\sum_{k=0}^{2n-1}e^{-2\pi\mathrm{i}jk/2n}\lambda_{\Xi_j,k}^{1/2}(y_{\mathrm{re},k}+\mathrm{i}y_{\mathrm{im},k}), \\
    \lambda_{l,j}=\frac{1}{\sqrt{2n}}\sum_{k=0}^{2n-1}e^{2\pi\mathrm{i}jk/2n}C_{\hat{\xi}_{l}}(\tau_k).
\end{gather}
\label{eq:circulant-embedding}
\end{subequations}

\begin{figure}[t]
    \centering
    \includegraphics[width=\textwidth]{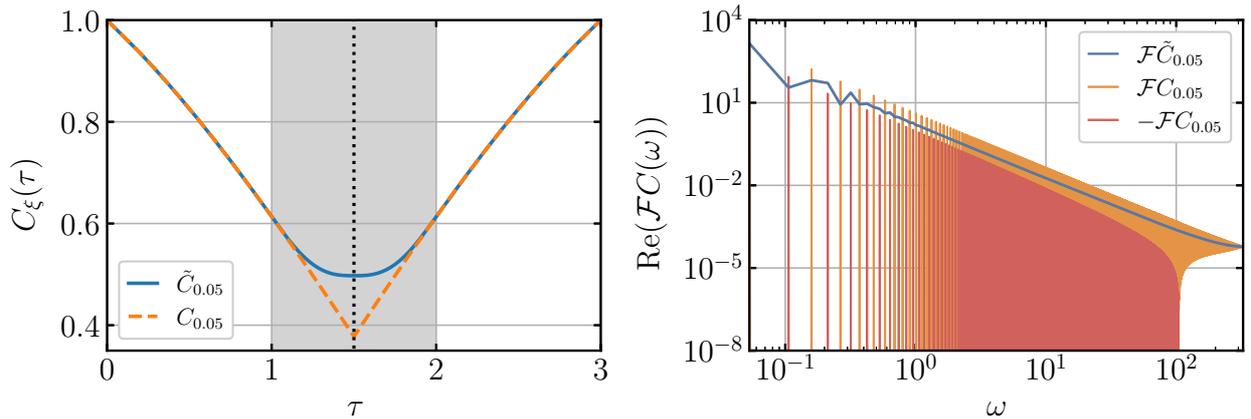}
    \caption{\emph{left:} Circulant embedding of the superstatistical covariance function with~$\tau\in(0,1.5)$,~$T=1$,~$\mathcal{T}=2$ and~$\xi=0.05$, with and without smoothing.
    The unmodified covariance function~$C_{0.05}(\tau)$ exhibits a pronounced discontinuity of the first derivative at the point of concatenation~$\tau_c=1.5$ leading to negative values of the Fourier transform, and thus an invalid covariance matrix with negative eigenvalues.
    This circumstance can be resolved by smoothly forcing the first derivative to zero on the transition region~$\tau\in(1,1.5)$ (indicated by the grey shaded region), according to equation~(\ref{eq:zero-derivative-circulant-embedding}), leading to the modified covariance function~$\tilde{C}(\tau)$.\\
    \emph{right:} Real parts of the discrete Fourier transforms of the circulant embeddings of~$C(\tau)$ and~$\tilde{C}(\tau)$.
    In the unmodified case, the Fourier transform oscillates strongly between positive and negative value. This behaviour is absent in the modified case.}
    \label{fig:zero-derivative-circulant-embedding}
\end{figure}

The covariance functions used in practice such as equation~(\ref{eq:matern}) are however not periodic.
In order to apply the Fourier algorithm, we use the \emph{circulant embedding} technique to construct periodic covariance matrices by concatenating the first row and its mirror image~$(C(\tau_0),\cdots,C(\tau_{n-1}))\in\mathbb{R}^{n}\to(C(\tau_0),\cdots,C(\tau_{n-1}),C(\tau_{n-2})\cdots,C(\tau_{1}))\in\mathbb{R}^{2n-2}$.
From the resulting samples of length~$2n-2$, only the first~$n$ elements are taken to remove the artificial periodicity again.
This increased memory requirement by a factor of 2 is more than compensated by the performance gain of the FFT algorithm.
However, circulant embedding must be done with care, because a too slow decay on the considered domain may lead to a pronounced discontinuity of the first derivative at the point of concatenation and negative eigenvalues of the circulant embedded matrix.
This problem could be circumvented by increasing the domain over which~$C(\tau)$ is sampled until sufficient decay is reached. Nonetheless, this approach might lead to critically increased memory requirements if~$C(\tau)$ decays too slowly.
Since this is the case with the superstatistical covariance function~$C_\xi(\tau)$ equation~(\ref{eq:superstat-covariance}) for small~$\xi\sim\mathcal{O}(10^{-2})$, we turn to a method loosely inspired by~\cite{helgason:2014} to smooth the discontinuity at the point of concatenation by smoothly forcing the derivative of the covariance function to zero on a transition region~$\tau\in(1,\tau_{\succ})$
\begin{equation}
    \Tilde{C}(\tau)=\begin{cases}
        C(\tau),\quad&\tau\in(0,1] \\
        C(1)+\int_1^\tau h\big(\frac{\tau'-1}{\tau_{\succ}-1}\big)C'(\tau')\dd{\tau'},\quad&\tau\in(1,\tau_{\succ})
    \end{cases}
    \label{eq:zero-derivative-circulant-embedding}
\end{equation}
where~$h(t)=2t^3-3t^2+1$ for~$t\in[0,1]$ is the first cubic Hermite spline basis function with~$h(0)=1$,~$h(1)=0$ and~$h'(0)=h'(1)=0$.
This is readily carried over to a discrete setting with~$C_j=C(\tau_j)$, where the transition region consists of~$m$ extra grid points
\begin{equation}
    \Tilde{C}_j=\begin{cases}
        C_j,\quad&j=0,\cdots,n-1 \\
        C_{n-1}+\sum_{k=n}^{j}h\big(\frac{k-n}{m-1}\big)(C_{k}-C_{k-1}),\quad&j=n,\cdots,n+m-1
    \end{cases}.
\end{equation}
The circulant embeddings of the raw and modified covariance functions~$C(\tau)$ and~$\tilde{C}(\tau)$ are compared in figure~(\ref{fig:zero-derivative-circulant-embedding}).
By taking only the first~$n$ points of samples generated with the circulant embedding of~$\Tilde{C}$, the effects of the transition region are removed.
The size~$\tau_\succ$ resp.~$m$ of the transition region is chosen such that all negative eigenvalues disappear. For our situation we found~$\tau_\succ=1.5$ to be an appropriate choice.

\section{Multiwavelets}
\label{sec:wavelets}
The previously described Fourier algorithm is an expansion of a Gaussian stochastic process with covariance function~$C(\tau)$ in terms of square-integrable orthogonal basis functions~$u(t)=\sum_{j=0}^{n}\hat{u}_j\varphi_j(t)$ with~$\int\varphi_j(t)\varphi_k(t)\dd{t}=\delta_{jk}$, where the coefficients~$\hat{u}_j$ are Gaussian random numbers with the transformed covariance matrix~$\hat{C}_{jk}=\iint C(|t-s|)\varphi_j(t)\varphi_k(s)\dd{t}\dd{s}$.
Specifically, in the Fourier case with~$C(\tau)$ periodic on~$\tau\in[0,T]$ we have~$\varphi_j(t)=\frac{1}{\sqrt{2\pi}}e^{2\pi\mathrm{i}jt/T}$ and~$\hat{C}_{jk}=\mathcal{F}\big[C\big(\frac{j}{T}\big)\big]\delta_{jk}$.
While the Fourier approach stands out for its algorithmic efficiency and diagonal covariance matrix, its requirement for uniform grids and linear scaling of memory with increasing resolution renders it impractical for local treatment of very large datasets.
Instead we consider basis functions derived from discrete wavelet transforms, i.e.~orthogonal functions with compact support, organized on a dyadic grid and with a prescribed number of vanishing moments. Such functions lead to representations of integral operators (which includes covariance functions) which are sparse to high precision and are thus a viable option for the expansion of stochastic processes~\cite{bcr:92,phoon:2004}.
Furthermore, the sparse covariance structure turn out to be helpful tools for efficient local interpolation of large datasets due to the vanishing moments and the decoupled scales of the dyadic grid.

In this work we employ a multiwavelet-valued generalization of the well-known Haar wavelet
\begin{equation}
    \psi(t)=\begin{cases}
        1,\quad&t\in[0,1/2) \\
        -1,\quad&t\in[1/2,1) \\
        0,\quad&\text{otherwise}
    \end{cases}
\end{equation}
due to Alpert~\cite{alpert-L2:1993}.
Such a multiwavelet of order~$q$ is a set of~$q$ square-integrable functions~$\{\psi_j\}_{j=0,\cdots,q-1}$ which are compactly supported on~$[0,1]$ and have at least~$q$ vanishing moments, i.e.~$\int_0^1t^m\psi_j(t)\dd{t}=0$ for~$m=0,\cdots,q-1$.
Note that for~$q=1$ the Haar wavelet is recovered.

\subsection{Multiresolution Analysis}
To motivate the construction of the multiwavelets of order~$q$, we review their associated Multiresolution Analysis (MRA) following~\cite{alpert-L2:1993} (see also~\cite{alpert-pde:2002}), i.e.~a decomposition of the space~$\mathcal{L}^2([0,1])$ of square-integrable functions on the interval~$[0,1]$ into increasingly accurate polynomial approximation and detail spaces.
On the coarsest scale\JAN{,}~$\mathcal{L}^2([0,1])$ is approximated by the space~$A^q_0$ of polynomials of degree less than~$q$. Given an orthogonal basis~$\phi_0,\cdots,\phi_{q-1}$ of~$A^q_0$ this approximation can be refined by by dilation and translation.
Specifically, at scale~$n>0$ there are~$2^nq$ functions
\begin{equation}
    \phi_p^{nk}(t)=2^{n/2}\phi_p(2^nt-k)
    \label{eq:scaling-refinement}
\end{equation}
for~$p=0,\cdots,q-1$ and~$k=0,\cdots,2^n-1$, compactly supported on the subintervals~$I_{nk}=[2^{-n}k,2^{-n}(k+1)]$, which span the finer approximation space~$A^q_n$.
Since any coarse function in~$A^q_{n_1}$ can be exactly expressed as a linear combination of finer functions in~$A^q_{n_2}$ with~$n_1<n_2$, we have a sequence of nested subspaces~$A_0^q\subset A_1^q\subset\cdots\subset A_n^q\subset\cdots$.
Furthermore, the detail lost by the approximation at scale~$n$ is captured by the orthogonal complement of~$A_n^q$ in~$A_{n+1}^q$ denoted by~$D_n^q$, i.e.~$A_n^q\oplus D_n^q=A_{n+1}^q$.
By iterating this expression, we arrive at the following decomposition of the~$n$-th approximation space:
\begin{equation}
    A_n^q=A_0^q\oplus D_0^q\oplus D_1^q\oplus\cdots\oplus D_{n-1}^q.
    \label{eq:subspaces-decomposition}
\end{equation}
A suitable ansatz for the basis functions of the detail space~$D^q_0$ are orthogonal piecewise polynomials~$\psi_0,\cdots,\psi_{q-1}$ of degree~$q-1$ defined on the subintervals~$I_{00}=[0,\frac{1}{2}]$ and~$I_{01}=[\frac{1}{2},{1}]$.
Note that due to~$A_0^q\perp D_0^q$, their first~$q$ moments naturally vanish.
The basis functions of~$n$-th order detail spaces are analogously to equation~(\ref{eq:scaling-refinement}) obtained by dilation and translation
\begin{equation}
    \psi_p^{nk}=2^{n/2}\psi_p(2^nt-k)
\end{equation}
for~$p=0,\cdots,q-1$ and~$k=0,\cdots,2^n-1$, compactly supported on the subintervals~$I_{nk}$ and satisfying orthogonality~$\int_0^1\psi^{nk}_i(t)\psi^{ml}_j(t)=\delta_{ij}\delta_{nm}\delta_{kl}$.
Then, due to equation~(\ref{eq:subspaces-decomposition}), the set 
\begin{equation}
    \{\phi_j\}_{j=0,\cdots,q-1}\cup\{\psi_p^{nk}\}_{j=0,\cdots,q-1;k=0,\cdots,2^n-1;n=0,\cdots,N-1}
    \label{eq:wavelet-basis}
\end{equation}
forms a basis of the approximation space~$A^q_N$.
It turns out that~$\lim_{N\to\infty}A^q_N$ is dense in~$\mathcal{L}^2([0,1])$, i.e.~for any square-integrable function~$f\in\mathcal{L}^2([0,1])$ there exists a sequence of functions~$\big(f_N\in A^q_N\big)_{N=0,1,\cdots}$ such that~$\lim_{N\to\infty}f_N=f$ uniformly.

The multiwavelets~$\psi_0,\cdots,\psi_{q-1}$ can be explicitly obtained in two ways:
\emph{(i)}~from a continuous point of view as piecewise polynomials of order~$q-1$ in~$\mathcal{L}^2([0,1])$, where the scaling functions are given by the Legendre polynomials shifted to~$[0,1]$,~$\phi_p(t)=\sqrt{2p-1}P_p(2t-1)$. Then the orthogonality and moment vanishing conditions can be translated in a linear system for the polynomial coefficients~\cite{alpert-L2:1993}.
And \emph{(ii)}~from a discrete point of view as vectors in a~$d$-dimensional vector space, where a~$d\times d$ orthogonal multiwavelet matrix~$\Psi$ is constructed as a solution to discrete versions of the orthogonality and moment vanishing conditions~\cite{alpert-linalg:1993,toolboxalperttransform}.
The matrix~$\Psi$ has the following form
\begin{equation}
    \Psi=\frac{1}{\sqrt{d}}\begin{pmatrix}
        \text{---}&\varphi_0(\mathbf{t})&\text{---} \\
        &\vdots&\\
        \text{---}&\varphi_d(\mathbf{t})&\text{---}
    \end{pmatrix},
\end{equation}
with row vectors~$\varphi_j(\mathbf{t})=(\varphi_j(t_0),\cdots,\varphi_j(t_{d-1}))$ and a~$d$-point discretization of the interval~$[0,1]$ given by~$\mathbf{t}=(t_0,\cdots,t_{d-1})$.
The scaling and multiwavelet functions are sorted as follows
\begin{equation}
    \varphi_j=\begin{cases}
        \phi_j\quad&\text{for }j<q, \\
        \psi_p^{nk}\quad&\text{for }j\ge q,\text{ with }j=p+q(2^n+k),
    \end{cases}
\end{equation}
where the multiwavelets are identifed by multiindices~$(n,k,p)$ with~$n$ varying the slowest and~$p$ varying the fastest.
The range of scales is~$n=0,\cdots,\log_2(d/2q)$ such that~$j=0,\cdots,d-1$. The multiwavelets on the last scale~$\log_2(d/2q)$ are supported on~$2q$ grid nodes, which is the smallest number of grid points for the discrete orthogonality and moment vanishing conditions to still hold.

\begin{figure}[t]
    \centering
    \includegraphics[width=.6\textwidth]{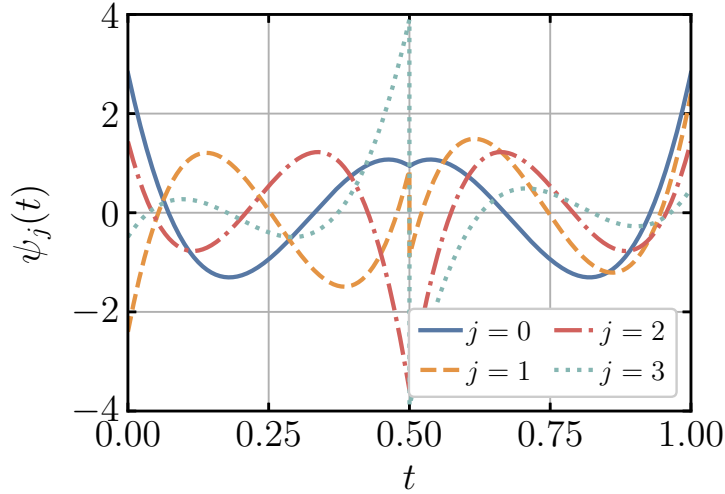}
    \caption{Alpert's multiwavelets of order~$q=4$.}
    \label{fig:multiwavelets}
\end{figure}

The multiwavelets of order~$q=4$ on scale~$n=0$ are plotted in figure~(\ref{fig:multiwavelets}).

\subsection{Multiwavelet expansion of stochastic process}
A Gaussian stochastic process with covariance function~$C(\tau)$, represented in the multiwavelet basis equation~(\ref{eq:wavelet-basis}) up to scale~$N$, reads
\begin{equation}
    u(t)=\sum_{j=0}^{M-1}\hat{u}_j\varphi_j(t)
    =\sum_{p=0}^{q-1}\hat{u}^\phi_p\phi_p(t)+\sum_{n=0}^{N-1}\sum_{k=0}^{2^n-1}\sum_{p=0}^{q-1}\hat{u}^\psi_{nkp}\psi^{nk}_p(t)
    \label{eq:wavelet-basis-process}
\end{equation}
with~$M=q2^N$ and
\begin{equation}
    (\hat{u}_j,\varphi_j)=\begin{cases}
        (\hat{u}^\phi_p, \phi_p) \quad&\text{with }j=p,\text{ for }j<q\\
        (\hat{u}^\psi_{nkp}, \psi^{nk}_p) \quad&\text{with }j=p+q(2^n+k),\text{ for }j\ge q
    \end{cases},
    \label{eq:dyadic-grid-linear-mapping}
\end{equation}
where the coefficients~$\hat{u}_j$ follow a discrete Gaussian process with the transformed covariance matrix
\begin{equation}
    \hat{C}_{jk}=\int_0^1\int_0^1C(|t-s|)\varphi_j(t)\varphi_k(s)\dd{t}\dd{s},
    \label{eq:covariance-transform}
\end{equation}
i.e.~the coefficient vector~$\hat{\mathbf{u}}=(\hat{u}_0,\cdots,\hat{u}_{M-1})=\hat{C}^{1/2}\mathbf{y}$ is given by the product of the matrix square root~$\hat{C}^{1/2}$ with a Gaussian white noise vector~$\mathbf{y}$.
Alternatively we consider a~$d$-point discretization of the interval~$[0,1]$ represented by grid points~$0\le t_0<\cdots<t_{d-1}\le 1$.
Then the multiwavelet basis functions given by equation~(\ref{eq:wavelet-basis}) are organized as rows of the orthogonal multiwavelet matrix~$\Psi$, which was introduced in the previous subsection, and the transformed covariance matrix is obtained from the similarity transform~$\hat{C}=\Psi\big(C(|t_j-t_k|)\big)_{j,k=0,\cdots,d-1}\Psi^\top$.
This leads to a matrix-vector product representation of the process equation~(\ref{eq:wavelet-basis-process})
\begin{gather}
    \mathbf{u}=\Psi^\top\hat{\mathbf{u}},\quad
    \hat{\mathbf{u}}=\hat{C}^{1/2}\mathbf{y}.
    \label{eq:discrete-wavelet-process}
\end{gather}
The two approaches correspond to each other with~$M=d$,~$N=\log_2(d/2q)+1$ and~$\varphi_j(t_k)=\sqrt{d}\Psi_{jk}$.

\begin{figure}[t]
    \centering
    \includegraphics[width=.8\textwidth]{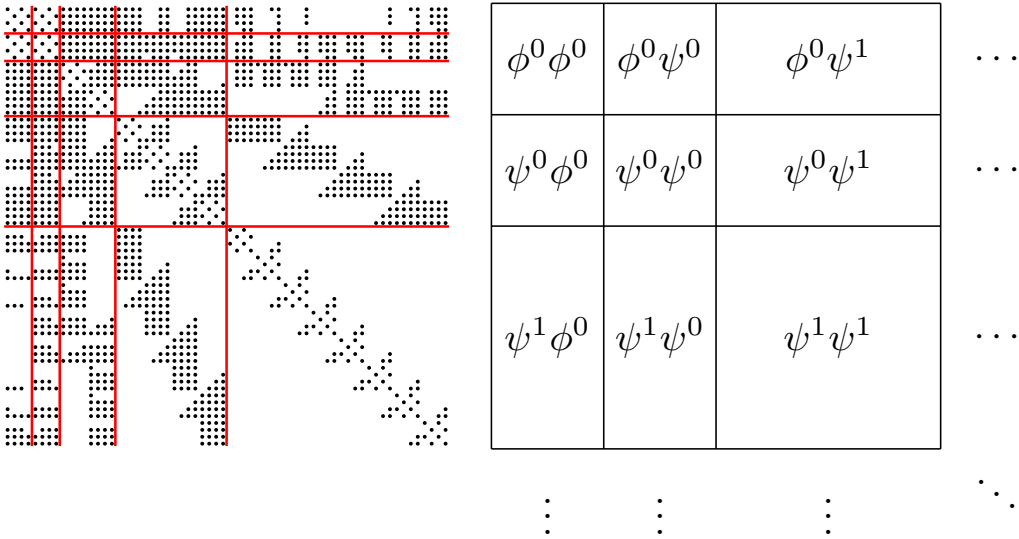}
    \caption{
    \emph{left:} The transformed covariance matrix~$\hat{C}$ with size~$M=64$, threshold~$\epsilon=5\times 10^{-5}$, non-zero elements colored black and blocks of scale pairs~$(n_1,n_2)$ highlighted by red lines.
    \emph{right:} Diagram of the scale pair block structure. Blocks denoted by~$\varphi_1^{n_1}\varphi_2^{n_2}$ contain correlations between scaling or wavelet functions~$\varphi_1,\varphi_2\in\{\phi,\psi\}$ at a fixed scale pair~$(n_1,n_2)$.}
    \label{fig:sparse-diagram}
\end{figure}

\begin{figure}[t]
    \centering
    \includegraphics[width=.35\textwidth]{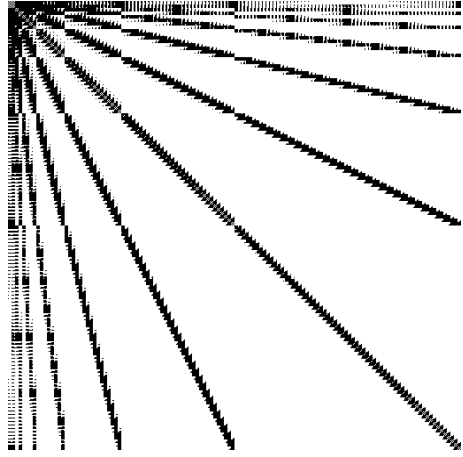}
    \caption{Example of the transformed covariance matrix~$\hat{C}$ with size~$M=512$ and threshold~$\epsilon=10^{-7}$, resulting in 12.9\% non-zero entries.}
    \label{fig:sparse-example}
\end{figure}

The transformed covariance matrix~$\hat{C}$ in the multiwavelet basis equation~(\ref{eq:wavelet-basis}) is sparse to high precision, i.e.~by setting all entries below some threshold~$\epsilon$ to zero, the number of non-zero elements is bounded by~$\mathcal{O}(M\log M)$ at the cost of a small controlled error~\cite{alpert-linalg:1993}.
Furthermore,~$\hat{C}$ exhibits a distinct block structure related to the multiresolution analysis, as illustrated in figures~(\ref{fig:sparse-diagram}) and~(\ref{fig:sparse-example}).
Specifically, the matrix is subdivided into blocks of sizes~$2^{n_1}q\times2^{n_2}q$ associated with scale pairs~$(n_1,n_2)$, where blocks on the main diagonal with~$n_1=n_2$ describe the interaction of wavelets on the same scale and off-diagonal blocks with~$n_1\neq n_2$ describe interactions of hierarchically separated wavelets.
While low-order blocks are rather dense, higher-order blocks become very sparse with non-zero elements only along distinct bands, which expresses spatially localized interaction among wavelets due to their vanishing moments.
We exploit this block-wise sparse structure when computing~$\hat{C}$ in the continuous case according to equation~(\ref{eq:covariance-transform}) to avoid computing~$M^2$ integrals. Specifically, the computation is done block-wise, where it starts at the main band of each block and walks outwards until a certain number of entries fall below the threshold~$\epsilon$.
These observations are valid for any covariance matrix obtained from a sufficiently regular covariance function.
Additionally, exploiting symmetry and stationarity of~$C(\tau)$ further reduce the number of integrals to be computed.

This sparse representation renders the process equation~(\ref{eq:formal-sampling}) applicable to large domains and higher spatial dimensions, which are typically inaccessible, due to readily available algorithms for sparse matrices~\cite{saad:2003,chow:2014}.
While the sparse structure in the multiwavelet representation is more complicated than the diagonal structure in Fourier representation, the latter is not subject to the requirement of uniform grids and naturally permits a localized description of the stochastic process.
In case of a localized description, a reduced covariance matrix can be constructed which takes only the relevant wavelets into account and thus enablesvery high resolutions locally.

\section{Stochastic interpolation in multiwavelet space}
\label{sec:interp}
Suppose that we want to interpolate a vector of measurements~$\mathbf{U}_{\mathbf{t}}$ sparsely sampled at times~$\mathbf{t}=(t_0,\cdots,t_{m-1})$ with a stochastic process realization~$\Tilde{\mathbf{u}}_{\mathbf{s}}$ at times~$\mathbf{s}=(s_0,\cdots,s_{n-1})$ with~$\mathbf{t}\subset\mathbf{s}$, such that~$\Tilde{\mathbf{u}}_{\mathbf{t}}=\mathbf{U}_{\mathbf{t}}$,
where the vector-valued indices refer to the times at which the time series are evaluated.
We assume that the measurement time series can be modeled by a Gaussian process with zero mean and covariance matrix~$\Sigma$.
For this purpose, Friedrich et al.~\cite{friedrich:2020} proposed the following interpolation formula, which reads in the notation of this paper
\begin{equation}
    \Tilde{\mathbf{u}}_{\mathbf{s}}=\mathbf{u}_{\mathbf{s}}+\Sigma_{\mathbf{st}}\Sigma_{\mathbf{tt}}^{-1}(\mathbf{U}_{\mathbf{t}}-\mathbf{u}_\mathbf{t}),
    \label{eq:bridge-equation}
\end{equation}
where~$\mathbf{u}$ (respectively evaluated at~$\mathbf{s}$ and~$\mathbf{t}$) denotes an unconstrained sample from a Gaussian process with zero mean and the covariance matrix~$\Sigma$.
The interpolating process~$\Tilde{\mathbf{u}}$ is also Gaussian with mean and covariance given by
\begin{subequations}
\begin{gather}
    \langle\Tilde{\mathbf{u}}\rangle=\Sigma_{\mathbf{st}}\Sigma_{\mathbf{tt}}^{-1}\mathbf{U}_{\mathbf{t}}, \\
    \langle\Tilde{\mathbf{u}}\Tilde{\mathbf{u}}^\top\rangle=\Sigma_{\mathbf{ss}}-\Sigma_{\mathbf{st}}(\Sigma_{\mathbf{tt}}^{-1}-(\Sigma_{\mathbf{tt}}^{-1}\mathbf{U})(\Sigma_{\mathbf{tt}}^{-1}\mathbf{U})^\top)\Sigma_{\mathbf{ts}},
\end{gather}
\end{subequations}
where the vector-valued indices refer to the times at which the covariance function is evaluated, e.g.~$(\Sigma_{\mathbf{ts}})_{jk}=C(t_j,s_k)$ for~$j=0,\cdots,m-1$ and~$k=0,\cdots,n-1$.
This approach utilizes unconstrained samples~$\mathbf{u}$, which is preferable compared to standard Gaussian process regression~\cite{rasmussen:2006} when an efficient unconstrained sampling algorithm such as circulant embedding is readily in place.

Since the multiwavelet coefficients~$\hat{\mathbf{u}}$ follow a Gaussian process with covariance matrix~$\hat{C}$, the interpolation formula can be straight-forwardly applied in multiwavelet space.
Specifically, we recall the multiwavelet expansion of the unconstrained process~$u(t)=\sum_{j=0}^{M-1}\hat{u}_j\varphi_j(t)$ and we suppose that~$\mu<M$ ``measured'' coefficients~$\hat{U}_{\gamma(j)}$ are given at indices~$j\in J\subset\{0,\cdots,M-1\}$ and~$\gamma:\{0,\cdots,M-1\}\to\{0,\cdots,\mu-1\}$ surjectively maps these coefficients to their respective positions on the full dyadic grid.
Then we want to find the multiwavelet expansion of the interpolating process~$\Tilde{u}(t)=\sum_{j=0}^{M-1}\hat{v}_j\varphi_j(t)$, such that~$\hat{v}_j=\hat{U}_{\gamma(j)}$ for all~$j\in J$.
To this end we denote the set of indices of the desired coefficients~$\hat{v}_j$ by~$K\subseteq\{0,\cdots,M-1\}$ with~$J\subset K$ and write equation~(\ref{eq:bridge-equation}) as
\begin{equation}
    \hat{\mathbf{v}}_K=\hat{\mathbf{u}}_K+\hat{C}_{KJ}\hat{C}_{JJ}^{-1}\big(\hat{\mathbf{U}}_{\gamma(J)}-\hat{\mathbf{u}}_J\big),
    \label{eq:wavelet-bridge-equation}
\end{equation}
where analogously to above the set-valued indices refer to the indices at which the quantities are evaluated.

\subsection{Affine Subspace Projection}
The difficult part in the here proposed interpolation approach is to properly determine the coefficients~$\hat{\mathbf{U}}$ from a time series of measurements~$\mathbf{U}$ of length~$m$ at times~$\mathbf{t}=(t_0,\cdots,t_{m-1})$.
For a uniformly sampled time series the multiwavelet expansion is well-defined to scale~$\log_2(m/2q)$, i.e.~the coefficients are obtained by projecting the time series onto the associated multiwavelets according to~$\hat{u}_j=\int_0^1u(t)\varphi_j(t)\dd{t}$,
which becomes in the discrete case a dot product~$\hat{U}_j=\frac{1}{\sqrt{m}}\varphi_j(\mathbf{t})^\top\mathbf{U}$ involving the~$j$-th row of the multiwavelet matrix~$\varphi_j(\mathbf{t})=\sqrt{m}\Psi_{j,:}=\big(\varphi_j(t_0),\cdots,\varphi_j(t_{m-1})\big)^\top$.
However, interpolation with high temporal resolution~$M>m$ also requires the consideration of higher scales~$\log_2(M/2q)>\log_2(m/2q)$, since a subset of multiwavelets on these unresolved scales carry a non-zero contribution to the measurements. Subsequently, their coefficients need to be determined for proper high resolution interpolation.

We express this situation formally by collecting the contributions of the~$\mu$ relevant multiwavelets up to scale~$\log_2(M/2q)$ in a~$\mu\times m$ matrix
\begin{equation}
    \Phi=\begin{pmatrix}
        \text{---}&\Tilde{\varphi}_0(\mathbf{t})&\text{---} \\
        &\vdots&\\
        \text{---}&\Tilde{\varphi}_{\mu-1}(\mathbf{t})&\text{---}
    \end{pmatrix},
\end{equation}
where each row has at least one non-zero entry and the matrix has naturally full column rank due to the dyadic structure of the wavelet basis (even if they are not fully resolved in time),
and writing each measurement as a weighted sum of the contributing multiwavelet as a linear system
\begin{equation}
    \mathbf{U}=\Phi^\top\hat{\mathbf{U}}.
    \label{eq:formal-linear-system}
\end{equation}
For equal resolution~$M=m$ we have the same number of coefficients~$\mu=m$ and for higher resolution~$M>m$ we have more coefficients than measurements~$\mu>m$,
thus in the latter case equation~(\ref{eq:formal-linear-system}) is an underdetermined linear system with infinitely many solutions for the coefficients~$\hat{\mathbf{U}}$.
Therefore, the key issue is to determine the correct solution considering that the coefficients follow a Gaussian process with covariance matrix~$\hat{C}_{JJ}$.
Making use of the Gaussian process property, we write~$\hat{\mathbf{U}}=\hat{C}^{1/2}_{JJ}\mathbf{y}$ with a Gaussian white noise vector~$\mathbf{y}$ and equation~(\ref{eq:formal-linear-system}) becomes
\begin{equation}
    \mathbf{U}=Z\mathbf{y}
    \label{eq:hyperplane-white-noise}
\end{equation}
with~$Z=\Phi^\top\hat{C}^{1/2}_{JJ}$.
Now this linear system can be interpreted as a Gaussian white noise vector constrained on the intersection of~$m$ affine hyperplanes, with normal vectors given by the row vectors of~$Z$ and distances from the origin given by the entries of~$\mathbf{U}$.
This intuition leads us to a solution of the form~$\mathbf{y}=\mathbf{y}_0+\mathbf{y}^\perp$, where~$\mathbf{y}_0=(Z^\top Z)^{-1}Z^\top\mathbf{U}$ is the minimum norm solution to equation~(\ref{eq:hyperplane-white-noise})
and~$\mathbf{y}^\perp$ is Gaussian white noise orthogonal to the image of~$Z$, i.e.~$Z\mathbf{y}^\perp=\mathbf{0}$, obtained by orthogonalizing unconstrained Gaussian white noise against the rows of~$Z$.
Both components~$\mathbf{y}_0$ and~$\mathbf{y}^\perp$ of the solution can be conveniently computed in terms of the QR decompostion of~$Z$.
The matrix~$Z$ is also sparse, making this procedure feasible for larger problems.

\subsection{Including Projected Coefficients}
One typically has the coarse wavelet coefficients~$\hat{\mathbf{U}}_0$ of the signal~$\mathbf{U}$ of length~$m$ available up to scale~$\log_2(m/q)$ due to a fast decomposition algorithm such as the one used in~\cite{alpert-pde:2002}.
For a desired resolution of~$M$ grid points,~$\log_2(M/m)$ scales are left unresolved, whose coefficients~$\hat{\mathbf{U}}_1$ can be determined with equation~(\ref{eq:hyperplane-white-noise}) and additionally taking advantage of the already available resolved coefficients.

To this end we write the square root of the reduced covariance matrix~$\hat{C}'=\hat{C}_{JJ}$ block-wise lower triangular
\begin{equation}
    S=\begin{pmatrix}
        S_{00}&0\\S_{10}&S_{11}
    \end{pmatrix},
\end{equation}
with~$S_{00}=\hat{C}'^{1/2}_{00}$,~$S_{10}=\hat{C}_{10}'\big(\hat{C}_{00}'^{1/2}\big)^{-\top}$ and~$S_{11}=Q^{1/2}$, where~$Q=\hat{C}'_{11}-\hat{C}'_{10}\hat{C}'_{00}\hat{C}'_{01}$ is the Schur complement of~$\hat{C}'_{00}$ in~$\hat{C}'$.
A quick calculation shows indeed~$SS^\top=\hat{C}'$.
%
We then divide the coefficients, wavelet contributions and the Gaussian white noise vector in resolved and unresolved components~$\hat{\mathbf{U}}=(\hat{\mathbf{U}}_0,\hat{\mathbf{U}}_1)^\top$,~$\Phi=(\Phi_0,\Phi_1)^\top$ and~$\mathbf{y}=(\mathbf{y}_0,\mathbf{y}_1)^\top$ in order to write the linear system equation~(\ref{eq:hyperplane-white-noise}) block-wise as
\begin{equation}
    \mathbf{U}=\begin{pmatrix}
        \Phi_0^\top&\Phi_1^\top
    \end{pmatrix}
    \begin{pmatrix}
        S_{00}&0\\S_{10}&S_{11}
    \end{pmatrix}
    \begin{pmatrix}
        \mathbf{y}_0\\\mathbf{y}_1
    \end{pmatrix}
    =\Phi_0^\top S_{00}\mathbf{y}_0+\Phi_1^\top(S_{10}\mathbf{y}_0+S_{11}\mathbf{y}_1).
\end{equation}
We note that the resolved Gaussian white noise vector is determined by the resolved coefficients according to~$\mathbf{y}_0=S_{00}^{-1}\hat{\mathbf{U}}_0$
and introduce the residual vector~$\mathbf{r}=\mathbf{U}-\big(\Phi_{0}^\top S_{00}+\Phi_1^\top S_{10}\big)\mathbf{y}_0$ to arrive again at an underdetermined linear system
\begin{equation}
    \mathbf{r}=\Phi_1^\top S_{11}\mathbf{y}_1,
    \label{eq:hyperplane-residual}
\end{equation}
which is solved for the unresolved Gaussian white noise vector~$\mathbf{y}_1$ in the same way as equation~(\ref{eq:hyperplane-white-noise}).
The resulting vector of ``measured'' coefficients, which can be plugged into the interpolation equation~(\ref{eq:wavelet-bridge-equation}), reads~$\hat{\mathbf{U}}=(\hat{\mathbf{U}}_0,S_{10}\mathbf{y}_0+S_{11}\mathbf{y}_1)^\top$.
In the following section, we illustrate the here proposed stochastic interpolation scheme on the basis of two examples, \emph{(i)}~the reconstruction of a subsampled signal and \emph{(ii)}~the local interpolation of a coarse signal.

\begin{figure}[t]
    \centering
    \includegraphics[width=\textwidth]{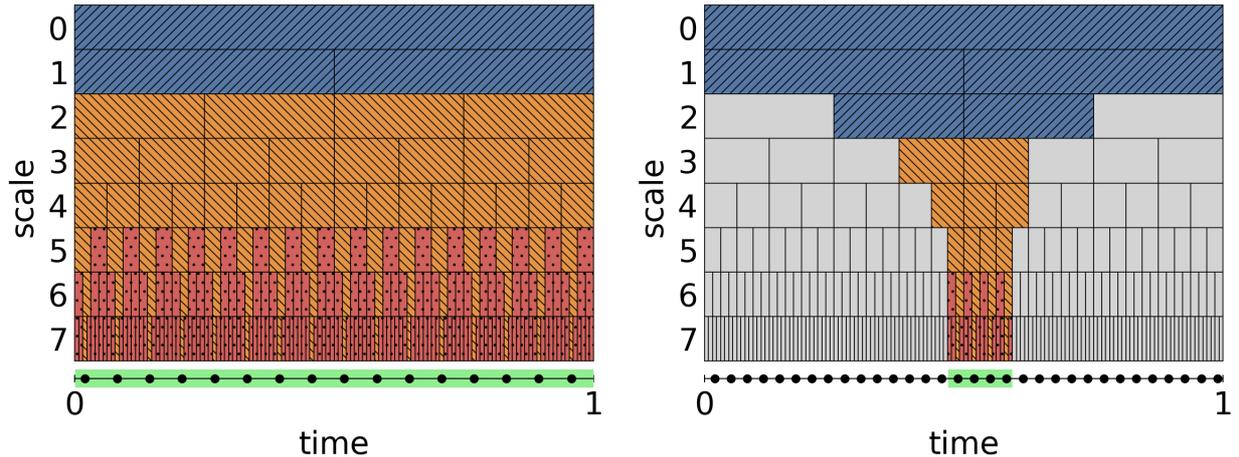}
    \caption{Contributing coefficients on the dyadic grid, highlighted according to their purpose.
    The region of interest for interpolation on the time axis is highlighted in green and the available measurements are marked by black dots.
    \emph{(a)}~Resolved coefficients, computed by deterministic wavelet decomposition, are marked in blue and forwardly hatched.
    \emph{(b)}~Unresolved coefficients contributing to the measurements, computed by the affine subspace projection, are marked in orange and backwardly hatched.
    \emph{(c)}~The remaining unresolved coefficients contributing to the interpolation, computed by the interpolation equation, are marked in red and dotted. \\
    \emph{left:} The region of interest is the full interval, reconstruction of an underlying signal.
    \emph{right:} The region of interest is a local subinterval, enabling efficient local high resolution.}
    \label{fig:dyadic-grid}
\end{figure}

\subsection{Examples}
We present two examples as technical tests of the interpolation procedure in multiwavelet space: 
\emph{(i)}~full reconstruction of a true signal given coarse measurements obtained through subsampling, and \emph{(ii)}~local interpolation with high resolution of a coarse signal.
Both cases follow the same scheme, as illustrated by the diagrams of dyadic grids in figure~(\ref{fig:dyadic-grid}):
Firstly, define the region of interest
over which to interpolate.
Secondly, collect the measurements which lie in this region of interest.
And thirdly, determined the multiwavelet coefficients which contribute to the region of interest.
These contributing coefficients are then divided into three disjoint sets: \emph{(a)}~resolved, \emph{(b)}~unresolved and contributing to the measurements in the region of interest, and \emph{(c)}~unresolved and contributing to the interpolation but not to the measurements.
The resolved coefficients are computed by a deterministic multiwavelet decomposition algorithm possibly applied to the entire signal, the unresolved measurement coefficients are computed by means of affine subspace projection, i.e.~solving equation~(\ref{eq:hyperplane-residual}), and the remaining unresolved interpolation coefficients are determined by the interpolation equation~(\ref{eq:wavelet-bridge-equation}).

\begin{figure}[tp]
    \centering
    \includegraphics[width=.5\textwidth]{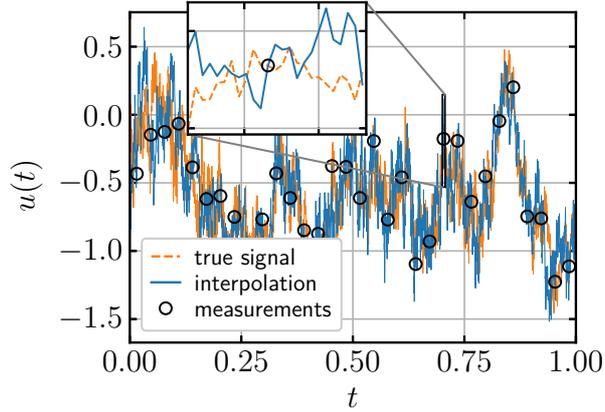}
    \caption{Reconstruction of a multifractal signal by means of multifractal stochastic interpolation of coarse measurements according to equation~(\ref{eq:wavelet-bridge-equation}) in combination with the mixture algorithm~(\ref{eq:formal-sampling}).
    An original signal of length 4096 synthesized with the mixture algorithm was downsampled by a factor 128 to length 32.
    This original signal is then reconstructed from these coarse ``measurements'' by synthesizing a conditioned realization of length 4096. \\
    These realizations and the ones in figure~(\ref{fig:superresolution}) were generated with the following parameters:~$\sigma=1$,~$H=1/3$,~$T=1$,~$\mathcal{T}=2$,~$T_{\mathrm{parameter}}=1$,~$A=0$,~$\mu=0.227$ and the~$\xi$-space discretization~$\log\hat{\xi}_j=-3+6j/99$ with~$j=0,\cdots,99$.
    }
    \label{fig:reconstruction}
\end{figure}

\begin{figure}[tp]
    \centering
    \includegraphics[width=\textwidth]{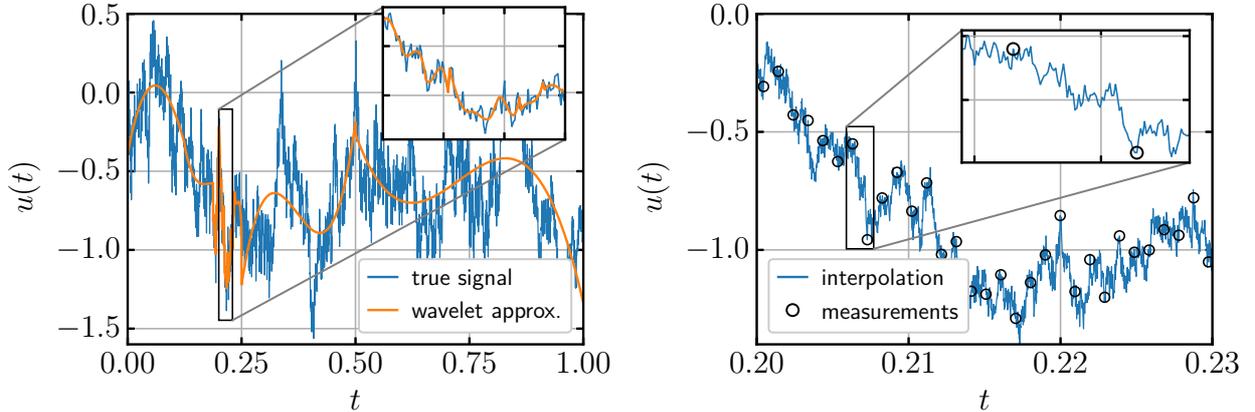}
    \caption{Local interpolation with high resolution of a synthetic multifractal signal with length 1024. The region of interest is the subinterval~$[0.2,0.23125]$ with 2048 grid points and the measurements are given directly by the true signal. This increases the resolution by a factor 64 and corresponds to an effective grid size of 65,536 nodes. \\
    \emph{left:} True signal and coarse multiwavelet approximation with adaptive accuracy. \emph{right:} Zoom of the region of interest. The measurements are the values of the true signal on the coarse grid.}
    \label{fig:superresolution}
\end{figure}

In case of reconstructing some underlying signal from a set of coarse measurements over the whole interval~$[0,1]$, all multiwavelet coefficients on the dyadic grid contribute, as illustrated in the left plot in figure~(\ref{fig:dyadic-grid}).
In contrast to this in case of local interpolation only a subset of coefficients are actually relevant, which are distributed with a funnel-like shape on the dyadic grid as seen in the right plot in figure~(\ref{fig:dyadic-grid}).
Note that the resolved coefficients computed by multiwavelet decomposition also take into account measurements outside of the region of interest and thus introduce correlation between the locally interpolated values and the coarse scales of the entire signal.
Example of both situations are depicted in figures~(\ref{fig:reconstruction}) and~(\ref{fig:superresolution}).

The appeal of the algorithm for local interpolation comes from the observation that the number of contributing coefficients (and therewith the size of the reduced covariance matrix) grows only logarithmically with the effective interpolation grid size.
Specifically, if~$r$ equidistant points with step size~$1/n$ are to be interpolated, the number of coefficients is bounded by~$r\log_2(n/2q)$ which scales as~$\mathcal{O}(\log_2n)$ for~$r\ll n$.
Additionally, the sparseness of the covariance matrix contributes to the efficiency of the algorithm.  

Finally, we note that this interpolation algorithm can be trivially combined with the superstatistical mixture algorithm equation~(\ref{eq:formal-sampling}) to give interpolating paths which exhibit log-normal multifractality.
Specifically, given a discretization of~$\xi$-space, where~$\xi$ is the mixture parameter of the covariance function, an interpolating path for each~$\xi_j$ is synthesized with identical underlying noise vectors.
Then each point in the region of interest gets assigned the corresponding value from one of these paths determined by a parameter process~$\xi(t)$, and the resulting path naturally interpolates the given measurements.
It is further possible to extract the coarse realization of the parameter process from a multifractal time series of measurements as described in~\cite{beck:2005} and use this result to condition the parameter process realization~$\xi(t)$ employed in the mixture algorithm.
However, this approach is left for future work.

\section{Results}
\label{sec:results}
We investigate the statistical properties of the mixture algorithm and the stochastic interpolation procedure in multiwavelet space by means of Monte-Carlo simulations.
Specifically, we carried out five simulations:
\begin{itemize}
    \item 
\emph{(i,ii)}~100,000 unconstrained realizations according to equation~(\ref{eq:formal-sampling}), where the underlying Gaussian processes were in \emph{(i)}~synthesized with the Fourier-based circulant embedding method given by equation~(\ref{eq:circulant-embedding}) and in \emph{(ii)}~with the Multiwavelet-based transformation of the covariance matrix according to equation~(\ref{eq:discrete-wavelet-process}).
    \item
\emph{(iii,iv)}~10,000 realizations reconstructing an underlying signal given coarse measurements of length 32, such as shown in figure~(\ref{fig:reconstruction}), where in \emph{(iii)}~a single reconstruction for each of 10,000 different measurement series were generated to verify that the reconstruction algorithm does not alter the statistics of the mixture process, and in \emph{(iv)}~10,000 reconstructions for a single measurement series were generated to investigate the behaviour in a more application-like setup.
    \item
\emph{(v)}~10,000 realizations locally interpolating a single given signal. The given signal is defined on the unit interval~$[0,1]$ with length 1024 and the interpolation is performed on the subinterval~$[0.2,0.2+2^{-5}]=[0.2,0.23125]$, which contains 32 point of the given signal, as shown in figure~(\ref{fig:superresolution}). Due to incorporating coarse wavelet coefficients, these local interpolations are conditioned on the entire signal. The interpolations consist of 2048 points, which corresponds to an upsampling factor of 64 and an effective resolution of 66,536 points on~$[0,1]$.
\end{itemize}
The signals used for conditioning in simulations \emph{(iii-v)}~were taken from the Fourier-based simulation \emph{(i)}~and appropriately downsampled.
The realizations of simulations \emph{(i-iv)}~have length 4096.
The common parameters of the covariance function~(\ref{eq:matern}) are~$\sigma=1$,~$T=1$ and~$H=1/3$, and the common parameters of the mixture variable~(\ref{eq:superstat-covariance}) are~$A=0$,~$\mu=0.227$
and~$\mathcal{T}=2$.
For the parameter process we chose the correlation time scale~$T_{\mathrm{parameter}}=1$ and the~$\xi$-space discretization as~$\log\hat{\xi}_j=-3+6j/99$ with~$j=0,\cdots,99$.

\begin{figure}[tp]
    \centering
    \includegraphics[width=.6\textwidth]{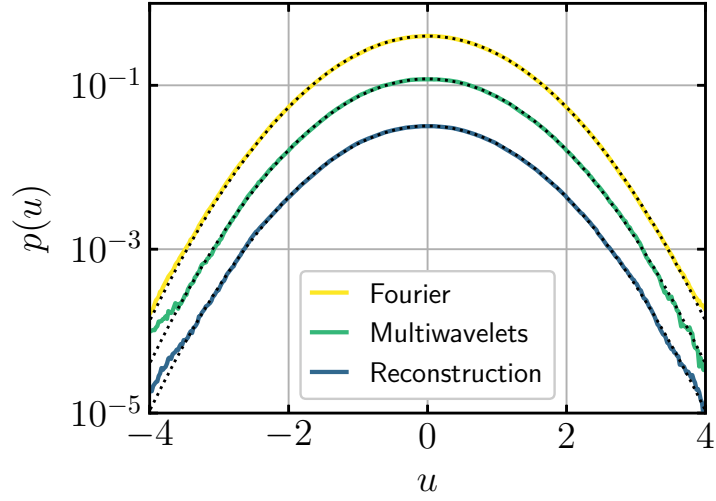}
    \caption{One-point distributions to confirm one-point Gaussianity according to the stationarity of the mixture process, as obtained from the multiwavelet-based sampling algorithm in the unconstrained, reconstruction and local interpolation cases.
    Gaussian distributions are given for reference as black dotted lines.
    The graphs are shifted vertically for clarity.}
    \label{fig:1pt}
\end{figure}

\begin{figure}[tp]
    \centering
    \includegraphics[width=.6\textwidth]{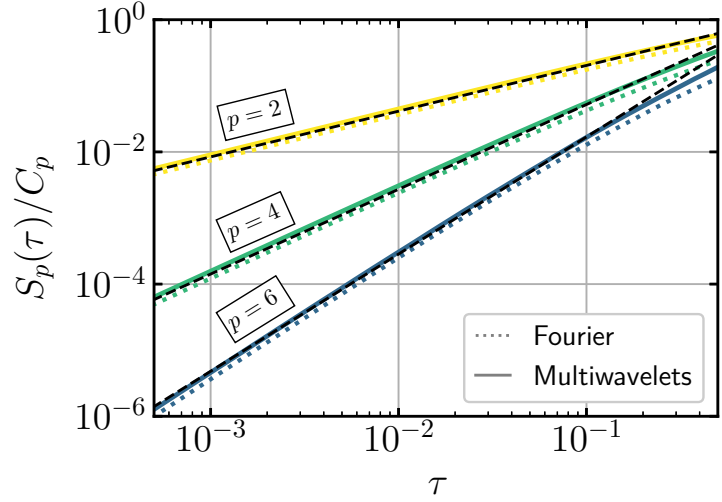}
    \caption{Structure functions of order~$p=2,4,6$ of the unconstrained Fourier- and multiwavelet-based simulations.
    The scaling laws~$S_p(\tau)=C_p\tau^{\zeta_p}$ are given for reference as black dashed lines with log-normal scaling exponents~$\zeta_p=\frac{p}{3}-\frac{\mu}{18}(p^2-3p)$ and constant factor~$C_p=(p-1)!!\mathcal{T}^{\frac{\mu}{18}(p^2-3p)}$ according to equation~(\ref{eq:Sp-constant-factor}).}
    \label{fig:Spu}
\end{figure}

\begin{figure}[tp]
    \centering
    \includegraphics[width=.6\textwidth]{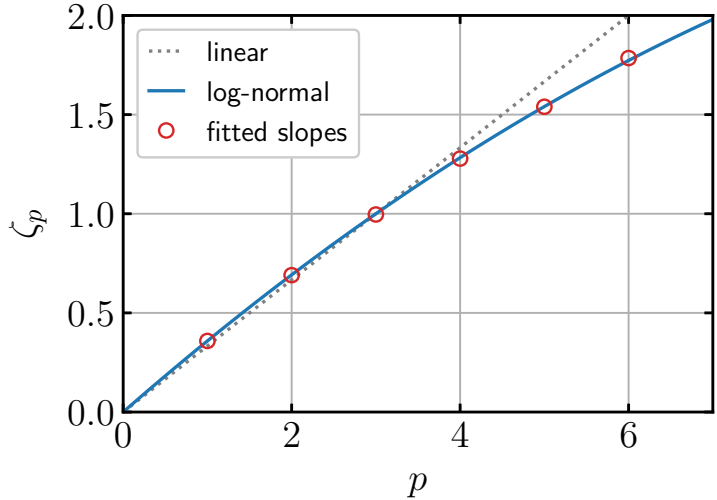}
    \caption{Scaling exponents~$\zeta_p$ for orders~$p=1,\cdots,6$, as obtained by linear fits of~$\log \tilde{S}_p(\tau)=\zeta_p\log\tau+\log C_p$, with~$\tilde{S}_p(\tau)=\langle|v_\tau|^p\rangle$ to account for odd orders~$p$.
    Only the exponents from the Fourier-based algorithm are shown.
    The linear scaling~$\zeta_p=p/3$ and the log-normal scaling~$\zeta_p=p/3-\frac{\mu}{18}(p^2-3p)$ are given for reference.}
    \label{fig:zeta}
\end{figure}

The one-point statistics for simulations \emph{(i-iii)}~shown in figure~(\ref{fig:1pt}) verify that the marginal distribution of mixture process and reconstruction algorithm is Gaussian, according to the stationarity of~$C_\xi(\tau)$.
The two-point statistics for the unconstrained simulations \emph{(i,ii)}~are investigated through the structure functions.~$S_p(\tau)$ is plotted for orders~$p=2,4,6$ against time lag~$\tau$ in figure~(\ref{fig:Spu}) and the results of both Fourier- and multiwavelet-base algorithms agree very well with the log-normal scaling laws on scales~$\tau<2\times 10^{-1}$.
Additionally, the scaling exponents were determined for orders~$p=1,\cdots,6$ by linear fits of~$\log\tilde{S}_p(\tau)=\zeta_p\log\tau+\log C_p$ with~$\tilde{S}_p(\tau)=\langle|v_\tau|^p\rangle$ and plotted in figure~(\ref{fig:zeta}).
The agreement of one-point and two-point statistics between Fourier- and multiwavelet-based algorithms sufficiently validates the multiwavelet-based approach.

\begin{figure}[tp]
    \centering
    \includegraphics[width=.6\textwidth]{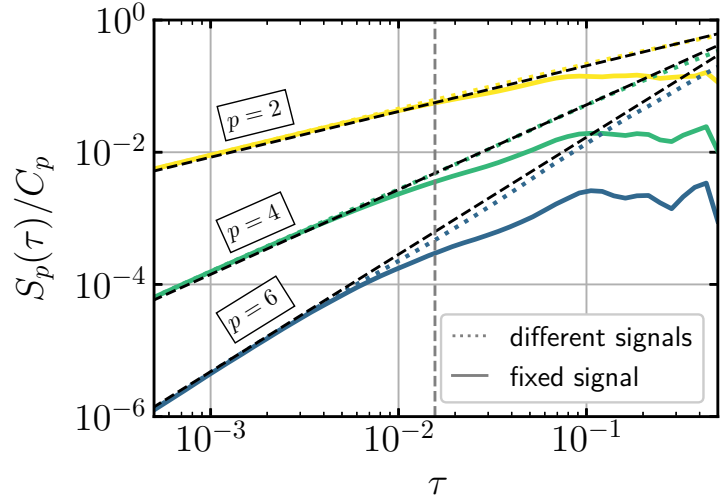}
    \caption{Structure functions of orders~$p=2,4,6$ of the reconstruction algorithm.
    The smallest scale of the coarse measurements~$\tau_{\mathrm{coarse}}=1/32$ is indicated by a vertical dashed line
    The log-normal scaling laws are again given for reference as black dashed lines.}
    \label{fig:Spc}
\end{figure}

The structure functions of orders~$p=2,4,6$ of the reconstruction simulations \emph{(iii,iv)}~are plotted in figure~(\ref{fig:Spc}).
For different underlying signals, the structure functions are indistinguishable from the unconstrained simulations, which validates the reconstruction algorithm.
For a single signal to reconstruct, like one would encounter in applications, we find log-normal scaling below the smallest scale of the coarse measurements~$\tau_{\mathrm{coarse}}$ and the expected deterioration on large scales, since only a single signal contributes to the statistics in that range.

\begin{figure}[tp]
    \centering
    \includegraphics[width=.6\textwidth]{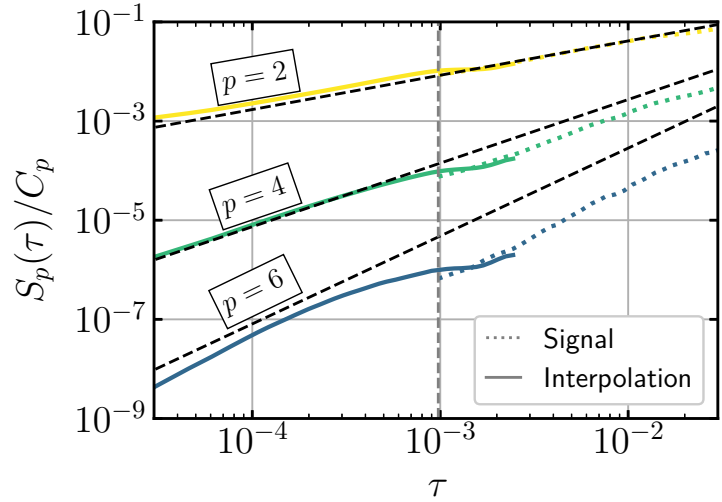}
    \caption{Structure functions of orders~$p=2,4,6$ of the local interpolation algorithm.
    The smallest scale of the given signal~$\tau_{\mathrm{coarse}}=1/1024$ is indicated by a vertical dashed line, and the log-normal scaling laws are again given for reference as black dashed lines.
    Note the transition between the structure functions of interpolation and signal.}
    \label{fig:Sps}
\end{figure}

Finally, the structure functions of orders~$p=2,4,6$ for the local interpolation simulation \emph{(v)}~and the given signal are plotted in figure~(\ref{fig:Sps}).
The small-scale local interpolations agree rather well with log-normal scaling and deterioration becomes again visible around the smallest scale of the given signal~$\tau_{\mathrm{coarse}}$.
We remark, that the structure functions of the interpolation transition almost smoothly into the structure functions of the signal.

\section{Discussion}
\label{sec:outlook}
In this work we presented a~$n$-point superstatistical model given for turbulent time series, which exhibit small-scale intermittency according to Kolmogorov's log-normal model with scaling exponents given by equation~(\ref{eq:lognormal-scaling}).
Specifically, we introduced a sampling algorithm of a numerically efficient approximation of the superstatistical process based on a discretization of~$\xi$-space, where~$\xi(t)$ is the mixture parameter process.
This algorithm consists of generating an ensemble of Gaussian processes~$u(t,\xi)$ with different covariance functions parameterized by~$\xi$ and identical underlying noise, and a point-wise choice procedure~$u(t)\gets u(t,\xi(t))$ to obtain an intermittent realization~$u(t)$.

Since the components of the ensemble are Gaussian, readily available interpolation formulae for Gaussian processes can be straight-forwardly applied.
However, the sampling and conditioning of Gaussian processes involve matrix square root computations and linear system solves involving the covariance matrix, which scale qubically and quadratically, respectively.
In order to facilitate these computations, we introduced a multiwavelet-based representation which leads to a particular sparse covariance structure and thus enables efficient local interpolation with drastically reduced matrix sizes.

We verified numerically that the herein presented sampling and conditioning algorithms faithfully reproduce log-normal scaling for structure functions of orders~$p=2,4,6$.
Furthermore it should be noted, that the local interpolation algorithm efficiently generates local small-scale samples accurately conditioned on the large scales of the entire given dataset.
Therefore, this approach offers itself for local high-resolution interpolation of coarse or incomplete datasets as they naturally appear e.g. in astrophysical settings.

A conceptual problem with the presented method becomes apparent in higher-order structure functions with~$p>6$, where the scaling exponents begin to deviate noticeably from the log-normal law.
This deviation is rooted in a discrepancy between the derivation of the scaling exponents in the superstatistical framework and the point-wise choice procedure.
Specifically, in the derivation of the structure functions given by equation~(\ref{eq:structure-functions}) we argue that a large correlation time scale~$T_{\mathrm{parameter}}$ of the parameter process~$\xi(t)$ justifies~$\xi(t+\tau)\approx\xi(t)$ on small scales, which is, strictly speaking, only exact for~$T_{\mathrm{parameter}}\to\infty$, i.e.~$\xi(t)=\mathrm{const}$ for all~$t$.
While this leads to the correct scaling exponents for all orders in terms of an ensemble average, it fails for individual realizations, i.e.~one obtains linear scaling in terms of a time average.
To obtain intermittency with a time average, a time-varying parameter process~$\xi(t)$ is required, where the correlation time scale~$T_{\mathrm{parameter}}$ controls the payoff between exact ensemble-level scaling and approximate sample-level scaling.

To facilitate deeper understanding of the matter, a rigorous treatment of regularity conditions of the form~$\langle(u(t_j,\xi_j)-u(t_k,\xi_k))^p\rangle\le f_p(|t_j-t_k|,|\xi_j-\xi_k|)$ with~$\xi_j\neq\xi_k$ is required.
Such a result could be used to derive conditions on the discretization of~$\xi$-space given a specific temporal resolution, such that the resulting approximated process~$u(t,\xi(t))$ is sufficiently regular.

A prospect for future work is the generalization of the superstatistical model to mixture distributions other than the log-normal model, for example an (inverse)~$\chi^2$ model~\cite{beck:2005}, the log-Poisson model by She and Leveque~\cite{she:1994} or the Yakhot model~\cite{yakhot:2006}.
In this context, it has been shown recently that the scaling of structure functions implies a particular Kramers-Moyal expansion~\cite{friedrich2020generalized}. Therefore, a first step into a more general non-Gaussian multipoint statistics would be to assess whether such a Kramers-Moyal expansion can be solved by a superstatistics as given by equation~(\ref{eq:increment-distribution-ensemble}).

Finally, a generalization to higher dimensions and vector-valued processes is conceptually straight-forward as outlined in~\cite{friedrich2022arxiv}, and the sparse covariance structure in wavelet-representation is a promising starting point for an efficient algorithm.
The model must also respect the divergence-free condition and offer the possibility for anisotropic energy spectra.
Further potential improvements include skewness of the velocity increment distribution and the intricate alignment behaviour of the vorticity with the eigenvectors of the strain field.
Eventually, this work should culminate in a fast algorithm for the evaluation of a turbulent magnetic field at a given particle position for the purpose of propagation studies of cosmic rays.

\section*{Acknowledgements}
R.G.~and J.L.~acknowledge funding from the German Science Foundation DFG, within the Collaborative Research Center SFB1491 ``Cosmic Interacting Matters - From Source to Signal''.
In addition, R.G.~and J.L.~acknowledge fruitful discussions with Timo Schorlepp.
J.F. acknowledges funding by the German Federal Ministry for Economic Affairs and Energy
in the scope of the projects EMUwind (03EE2031A/C) as well as by the Humboldt Foundation within a Feodor-Lynen fellowship.

\begingroup
\sloppy
\printbibliography
\endgroup

\clearpage

\appendix 

\newcommand{\erf}{\mathrm{erf}}

\section{Derivation of the superstatistical Structure Functions}
\label{app:moments}
Let~$u(t,\xi(t))$ denote the continuous mixture process introduced in section \ref{sec:algo}, where~$u_\xi(t)=u(t,\xi)$ with fixed~$\xi$ denotes a Gaussian process with covariance function~$C_\xi(\tau)$ according to equation~(\ref{eq:superstat-covariance}).
In order to prove that the mixture process~$u(t,\xi(t))$ indeed exhibits log-normal multifractality, we compute the scale-dependent moments of the increment process~$v_\tau(t)=u(t+\tau,\xi(t+\tau))-u(t,\xi(t))$ over small scales~$\tau\ll T$, i.e.~its structure functions.
According to the requirement on the parameter process~$\xi(t)$ to have a large correlation time scale~$T_{\mathrm{parameter}}\gg T$, we assume on average~$\xi(t+\tau)\approx\xi(t)$ for sufficiently small~$\tau$, and can then write down the variance of the increments, i.e.~the second-order structure function
\begin{equation}
    S_{2,\xi}(\tau)=\langle(u_\xi(t+\tau)-u_\xi(t))^2\rangle=2C_\xi(0)-2C_\xi(\tau)\approx\sigma^2(\varepsilon_{\xi,\tau}\tau)^{2H},
\end{equation}
where we made use of the stationarity of~$u_\xi(t)$ and the small-scale asymptotics of~$C_\xi(\tau)$.
According to the superstatistical framework, the scale-dependent distribution of the increments of the mixture process is then obtained by averaging over all~$\xi$, weighted by a log-normal distribution
\begin{equation}
    p(v_\tau,\tau)=\frac{1}{2\pi}\int_0^\infty\frac{\dd{\xi}}{\xi}\frac{1}{\sqrt{S_{2,\xi}(\tau)}}\exp\left(-\frac{1}{2}[\log\xi]^2]\right)\exp\left(-\frac{v_\tau^2}{2S_{2,\xi}(\tau)^2}\right).
\end{equation}
We recall~$\varepsilon_{\xi,\tau}=\xi^{\sqrt{A+\mu\log\mathcal{T}/\tau}}\left(\tau/\mathcal{T}\right)^{\mu/2}$ and compute the even moments of this distribution analytically
\begin{align}
    \langle v_\tau^q\rangle
    &=\int_{-\infty}^{+\infty}\dd{v_\tau}v_\tau^qp(v_\tau,\tau) \nonumber\\
    &=\int_0^\infty\frac{\dd{\xi}}{\xi}\frac{1}{\sqrt{2\pi}}\exp\left(-\frac{1}{2}[\log\xi]^2\right)\sigma^{q}(q-1)!!\,(\varepsilon_{\xi,\tau}\tau)^{qH} \nonumber\\
    &=\frac{\sigma^{q}(q-1)!!}{\sqrt{2\pi}}\left(\frac{\tau}{\mathcal{T}}\right)^{qH\mu/2}\tau^{qH}\times\nonumber\\
    &\qquad\times\int_0^\infty\frac{\dd{\xi}}{\xi}\exp\left(-\frac{1}{2}[\log\xi]^2+qH\sqrt{A+\mu\log{\mathcal{T}}/{\tau}}\log\xi\right) \nonumber\\
    &=\frac{\sigma^{q}(q-1)!!}{\sqrt{2\pi}}\left(\frac{\tau}{\mathcal{T}}\right)^{qH\mu/2}\tau^{qH} \times\nonumber\\
    &\qquad\times\sqrt{\frac{\pi}{2}}\exp\left(\frac{q^2H^2}{2}\left(A+\mu\log{\mathcal{T}}/{\tau}\right)\right) 
    \left[\erf\left(\frac{qH}{\sqrt{2}}\sqrt{A+\mu\log{\mathcal{T}}/{\tau}}\right)+1\right] \nonumber\\
    &\approx\sigma^{q}(q-1)!!\,\exp\left(\frac{1}{2}q^2H^2A\right)
    \left(\frac{\tau}{\mathcal{T}}\right)^{qH\mu/2}\tau^{qH}\left(\frac{\tau}{\mathcal{T}}\right)^{-q^2H^2\mu/2} \nonumber \\
    &=C_q\tau^{qH+\frac{\mu}{2}(qH-q^2H^2)}.
    \label{eq:superstatstructurefuncs}
\end{align}
Here we employed the~$q$-th moment of a Gaussian random variable~$\langle X^q\rangle=\langle X^2\rangle^{q/2}(q-1)!!$ for even~$q$,~$\langle X^2\rangle=\sigma$,
the Gaussian integral~$\int_0^\infty\dd{x} e^{-x^2/2+bx}=\sqrt{\frac{\pi}{2}}e^{b^2/2}\left(\erf\left({b}/{\sqrt{2}}\right)+1\right)$, where~$\erf(x)=\frac{2}{\sqrt{\pi}}\int_0^xe^{-t^2}\dd{t}$ is the Gaussian error function,
and approximated the error function expression as 1, because its argument is sufficiently large for~$\tau\ll\mathcal{T}$.
The structure functions thus follow a power law with scaling exponent
\begin{equation}
    \zeta_q=qH+\frac{\mu}{2}(qH-q^2H^2),
\end{equation}
which coincides with Kolmogorov's log-normal model of intermittency according to equation~(\ref{eq:lognormal-scaling}) for~$H=1/3$,
and constant factor
\begin{equation}
    C_q=\sigma^q(q-1)!!\,\exp\left(\frac{1}{2}q^2H^2A\right)\mathcal{T}^{\frac{\mu}{2}(q^2H^2-qH)}.
    \label{eq:Sp-constant-factor}
\end{equation}

\end{document}